\documentclass[aps,twocolumn,floatfix,amsmath,amsfonts,longbibliography]{revtex4-2}

\usepackage{graphicx}
\usepackage{braket}
\usepackage{bm}
\usepackage{xcolor}
\usepackage{upgreek}
\usepackage{tabularx}
\usepackage[normalem]{ulem}
\usepackage[utf8]{inputenc}
\usepackage{amsmath}
\usepackage{amssymb}
\usepackage{todonotes}
\usepackage{color}
\usepackage{epsfig}
\usepackage{comment}
\usepackage{siunitx}
\renewcommand{\arraystretch}{2}
\newcommand{\bk}{\boldsymbol{k}}

\newcommand{\bq}{\boldsymbol{q}}
\newcommand{\bp}{\boldsymbol{p}}
\newcommand{\bP}{\boldsymbol{P}}
\newcommand{\bpi}{\boldsymbol{\pi}}
\newcommand{\bsi}{\boldsymbol{\sigma}}
\newcommand{\bv}{\boldsymbol{v}}
\newcommand{\br}{\boldsymbol{r}}

\newcommand{\bA}{\boldsymbol{A}}
\newcommand{\rmc}{\mathrm{c}}
\newcommand{\rmv}{\mathrm{v}}

\newcommand{\cuo}{Cu$_{\mathrm{2}}$O}

\newcommand{\ac}[1]{\multicolumn{1}{c|}{#1}}

\usepackage{hyperref}

\begin{document}
\title{Interseries dipole transitions from yellow to green excitons in cuprous oxide}
\author{Patric Rommel}
\email[Email: ]{patric.rommel@itp1.uni-stuttgart.de}
\author{Jörg Main}
\affiliation{Institut für Theoretische Physik 1, Universität
  Stuttgart, 70550 Stuttgart, Germany}
\author{Sjard Ole Krüger}
\author{Stefan Scheel}
\affiliation{Institut für Physik, Universität Rostock,
  Albert-Einstein-Straße 23-24, 18059 Rostock, Germany}

\date{\today}

\begin{abstract}
We study dipole interseries transitions between the yellow and green
exciton series in cuprous oxide including the complex valence band
structure.
To this end, we extend previous studies of the spectrum of complex
green exciton resonances 
[Phys.\ Rev.\ B \textbf{101}, 075208 (2020)] to optical transitions between 
different exciton states in addition to transitions from the crystal ground 
state. This allows us to augment the calculations on 
interseries transitions using a hydrogen-like model
[Phys.\ Rev.\ B \textbf{100}, 085201 (2019)] by a more
comprehensive treatment of the 
valence band structure.
\end{abstract}

\maketitle

\section{Introduction}
\label{sec:introduction}
Cuprous oxide has long been a very interesting system for the study of 
excitons. Indeed, it was in this material that excitons were first observed 
\cite{gross1952,Gross1956}, and in which bound states with principal quantum 
numbers up to $n=25$ have been detected \cite{GiantRydbergExcitons}. This 
abundance of known resonances allows for very precise tests of theoretical 
models, and to probe the influence of intricate details of the band structure on 
the formation of excitons.

Most of the work in the literature focuses on the yellow series, which
is formed by electrons in the lowest $\Gamma_6^+$ conduction band
and holes in the uppermost $\Gamma_7^+$ valence band 
\cite{ObservationHighAngularMomentumExcitons,ImpactValence,Elliot1961}.
The green excitons, on the other hand, are formed by holes in the
$\Gamma_8^+$ valence band~\cite{Grun1961,Nikitine1959,Nikitine1963}. For principal quantum numbers $n \geq 2$, they are 
located at energies above the band gap of the yellow excitons, and couple to 
the yellow continuum states. Thus, even without taking phonon coupling into 
account, the green excitons above the yellow band gap are no longer bound states with 
infinite lifetimes, but quasi-bound resonances with finite lifetimes.
Recently, 
the locations of the green exciton resonances have been calculated 
\cite{Rommel2020Green} using the complex-scaling
method \cite{MOISEYEV1998,Zielinski2020}.

Motivated by the aim to identify promising experimentally accessible dipole 
transitions for the coherent manipulation of Rydberg excitons 
\cite{PhysRevLett.117.133003,Khazali_2017} and the generation
of giant optical nonlinearities \cite{Walther2018}, 
interseries transitions between the yellow and green, respectively yellow and 
blue, exciton series have been investigated using a hydrogen-like model for the 
exciton interaction \cite{KruegerInterseries2019}.
The interseries transitions, i.e., those between different exciton series,
have the distinct advantage over intraseries transitions, i.e., transitions 
within a single exciton series, by providing a more accessible choice of 
interrogation wavelengths. To wit, transition wavelengths between adjacent 
Rydberg states within the same series scale as $n^3$ and quickly approach the 
millimeter range, whereas the wavelength limit for interseries transitions is 
set by the energy difference between bands, which are typically in the near- to 
mid IR.
In this article, we investigate interseries dipole transitions between the 
yellow and green exciton series, while taking into account the complex structure 
of the valence band \cite{Luttinger56,ImpactValence} as well as central-cell 
corrections 
\cite{frankevenexcitonseries,knox1963excitons,Uihlein1981,Froehlich1979,
KavoulakisBohrRadius}.
As our focus is not on optical transitions where the exciton is created 
from the crystal ground state, but rather on transitions between different 
exciton states, this requires an extensive modifications of the scheme for
calculating the oscillator strengths.

The article is organized as follows.
First, we present the numerical calculation of the relevant exciton
states using a unified Hamiltonian describing both the yellow and
green series in Sec.~\ref{sec:CalculationStates}.
Using the calculated eigenvalues and eigenvectors, we derive the
dipole transition matrix elements in Sec.~\ref{sec:DipoleTransitions}.
In Sec.~\ref{sec:Results}, we present and discuss our results on
interseries dipole transitions and absorption spectra.
We finish the paper with some concluding remarks in Sec.~\ref{sec:Conclusion}.

\section{The spectrum of yellow and green excitons}
\label{sec:CalculationStates}
In this section we briefly recapitulate and extend our technique for
calculating the bound yellow exciton states and the unbound
green exciton resonances by using a complete basis set and the
complex-coordinate rotation method.
For both the yellow and green series, we take the valence
band structure and central-cell corrections into account.
The knowledge of the precise states is the prerequisite for the
computation of interseries dipole transitions in
Secs.~\ref{sec:DipoleTransitions} and \ref{sec:Results}.

\subsection{Hamiltonian}
\label{subsec:Hamiltonian}
The description of excitons follows the line of arguments laid out in 
Ref.~\cite{Rommel2020Green}. For the investigation of interseries transitions 
between yellow and green exciton states, we use the unified description of the 
two series given by the Hamiltonian \cite{Rommel2020Green,
ImpactValence,Schoene2016,Luttinger56,frankevenexcitonseries}
\begin{equation}
  H = E_{\rm g}+\frac{\gamma_1'}{2m_{0}}\boldsymbol{p}^2+H_{\rm b}(\boldsymbol{p})
  -\frac{e^2}{4\pi\varepsilon_0\varepsilon|\boldsymbol{r}|}\ + H_{\mathrm{CCC}}\, ,
\label{eq:hamiltonian}
\end{equation}
with the valence-band corrections to the kinetic energy
\begin{align}
  H_{\rm b}(\boldsymbol{p}) &= H_{\rm SO}+\frac{1}{2\hbar^2m_0} \big\{4\hbar^2\gamma_2\boldsymbol{p}^2\label{eq:holeKinetic}
+2(\eta_1+2\eta_2)\boldsymbol{p}^2(\boldsymbol{I}\cdot\boldsymbol{S}_{\rm h})\nonumber\phantom{\frac{1}{2}}\\
&-6\gamma_2(p^2_{1}\boldsymbol{I}^2_1+{\rm c.p.})
-12\eta_2(p^2_{1}\boldsymbol{I}_1\boldsymbol{S}_{\rm h1}+{\rm c.p.})\nonumber\phantom{\frac{1}{2}}\\
&-12\gamma_3(\{p_{1},p_{2}\}\{\boldsymbol{I}_1,\boldsymbol{I}_2\}+{\rm c.p.})\nonumber\phantom{\frac{1}{2}}\\
\phantom{\frac{1}{2}} &-12\eta_3(\{p_{1},p_{2}\}(\boldsymbol{I}_1\boldsymbol{S}_{\rm h2}
         +\boldsymbol{I}_2\boldsymbol{S}_{\rm h1})+{\rm c.p.})\big\} \, 
\end{align}
and the central-cell corrections $H_{\mathrm{CCC}}$ discussed below in
Sec.~\ref{sec:central-cell_corrections}.
Here we use center-of-mass coordinates \cite{ImpactValence,Schmelcher1992},
\begin{align}
  \boldsymbol{r}&= \boldsymbol{r}_{\rm e}-\boldsymbol{r}_{\rm h}\, ,\quad
  \boldsymbol{R}=\frac{m_{\rm h}\boldsymbol{r}_{\rm h}+m_{\rm e}\boldsymbol{r}_{\rm e}}{m_{\rm h}+m_{\rm e}}\, ,\nonumber\\
  \boldsymbol{P}&=\hbar\boldsymbol{K}= \boldsymbol{p}_{\rm e}+\boldsymbol{p}_{\rm h}\, ,\quad
  \boldsymbol{p}=\hbar\boldsymbol{k}=\frac{m_{\rm h}\boldsymbol{p}_{\rm e}-m_{\rm e}\boldsymbol{p}_{\rm h}}{m_{\rm h}+m_{\rm e}} \, ,
  \label{eq:COMcoordinates}
\end{align}
with $\boldsymbol{r}_\mathrm{e}$ and $\boldsymbol{r}_\mathrm{h}$ the
electron and hole positions, and $\boldsymbol{p}_\mathrm{e}$ and
$\boldsymbol{p}_\mathrm{h}$ their corresponding momenta.
The center-of-mass momentum is set to $\boldsymbol{P} = 0$.
We use the effective and free electron masses $m_{\rm e}$ and $m_0$,
respectively, the symmetrized product $\{a,b\}=\frac{1}{2}(ab+ba)$,
the Luttinger parameters $\gamma_j$ and $\eta_j$, and
$\gamma_1'=\gamma_1+m_0/m_{\mathrm{e}}$,
$E_\mathrm{g}$ is the gap energy, $\varepsilon$ the dielectric
constant, and c.p.\ denotes cyclic permutation.
The band structure Hamiltonian~\eqref{eq:holeKinetic} necessitates the introduction of
the hole spin $\boldsymbol{S}_\mathrm{h}$ and the quasispin $\boldsymbol{I}$, the latter allowing
for a convenient description of the degenerate $\Gamma_5^+$ valence bands.
The spin-orbit coupling,
\begin{equation}
  H_{\rm SO}=\frac{2}{3}\Delta
  \left(1+\frac{1}{\hbar^2}\boldsymbol{I}\cdot\boldsymbol{S}_{\rm h}\right)\,,
  \label{eq:SOcoupling}
\end{equation}
leads to an energetic splitting $\Delta$ of the valence bands into the higher lying
$\Gamma_7^+$ bands associated with the yellow exciton series and the
lower lying $\Gamma_8^+$ bands associated with the green exciton series.
The material parameters of cuprous oxide used in our calculations are listed
in Table~\ref{tab:MaterialParamters}.

\begin{table}[b]
\renewcommand{\arraystretch}{1.2}
     \centering
     \caption{Material parameters of Cu$_2$O used in the calculations.}
     \begin{tabular}{l|l c}
     \hline
       Energy gap  & $E_{\rm g}=2.17208\,$eV & \cite{GiantRydbergExcitons}\\
       Spin-orbit coupling       & $\Delta=0.131\,$eV &\cite{SchoeneLuttinger}\\
       Effective electron mass  & $m_{\rm e}=0.99\,m_0$ & \cite{HodbyEffectiveMasses} \\
       Effective hole mass  & $m_{\rm h}=0.58\,m_0$ & \cite{HodbyEffectiveMasses} \\
       Luttinger parameters & $\gamma_1=1.76$&\cite{SchoeneLuttinger}\\
       ~~& $\gamma_2=0.7532$&\cite{SchoeneLuttinger}\\
       ~~& $\gamma_3=-0.3668$&\cite{SchoeneLuttinger}\\
        ~~& $\eta_1=-0.020$&\cite{SchoeneLuttinger}\\
        ~~& $\eta_2=-0.0037$&\cite{SchoeneLuttinger}\\
        ~~& $\eta_3=-0.0337$&\cite{SchoeneLuttinger}\\
       Exchange interaction & $J_0 = 0.792\,\mathrm{eV}$ & \cite{frankevenexcitonseries} \\
       Short distance correction & $V_0 = 0.539\,\mathrm{eV}$ & \cite{frankevenexcitonseries}\\
       Lattice constant & $a=0.42696\,\mathrm{nm}$ & \cite{Swanson1953}\tabularnewline
       Dielectric constants & $\varepsilon_{\mathrm{s}1}=\varepsilon=7.5$ & \cite{LandoltBornstein1998DielectricConstant}\tabularnewline
                            & $\varepsilon_{\mathrm{b}1}=\varepsilon_{\mathrm{s}2}=7.11$ & \cite{LandoltBornstein1998DielectricConstant}\tabularnewline
                            & $\varepsilon_{\mathrm{b}2}=6.46$ & \cite{LandoltBornstein1998DielectricConstant}\tabularnewline
       Energy of $\Gamma_{4}^{-}$-LO phonons & $\hbar\omega_{\mathrm{LO1}}=18.7\,\mathrm{meV}$ & \cite{KavoulakisBohrRadius}\tabularnewline
                            & $\hbar\omega_{\mathrm{LO2}}=87\,\mathrm{meV}$ & \cite{KavoulakisBohrRadius}\tabularnewline

     \hline
     \end{tabular}
\label{tab:MaterialParamters}
\end{table}

\subsection{Complex-coordinate rotation}
\begin{figure}
\includegraphics[width=\columnwidth]{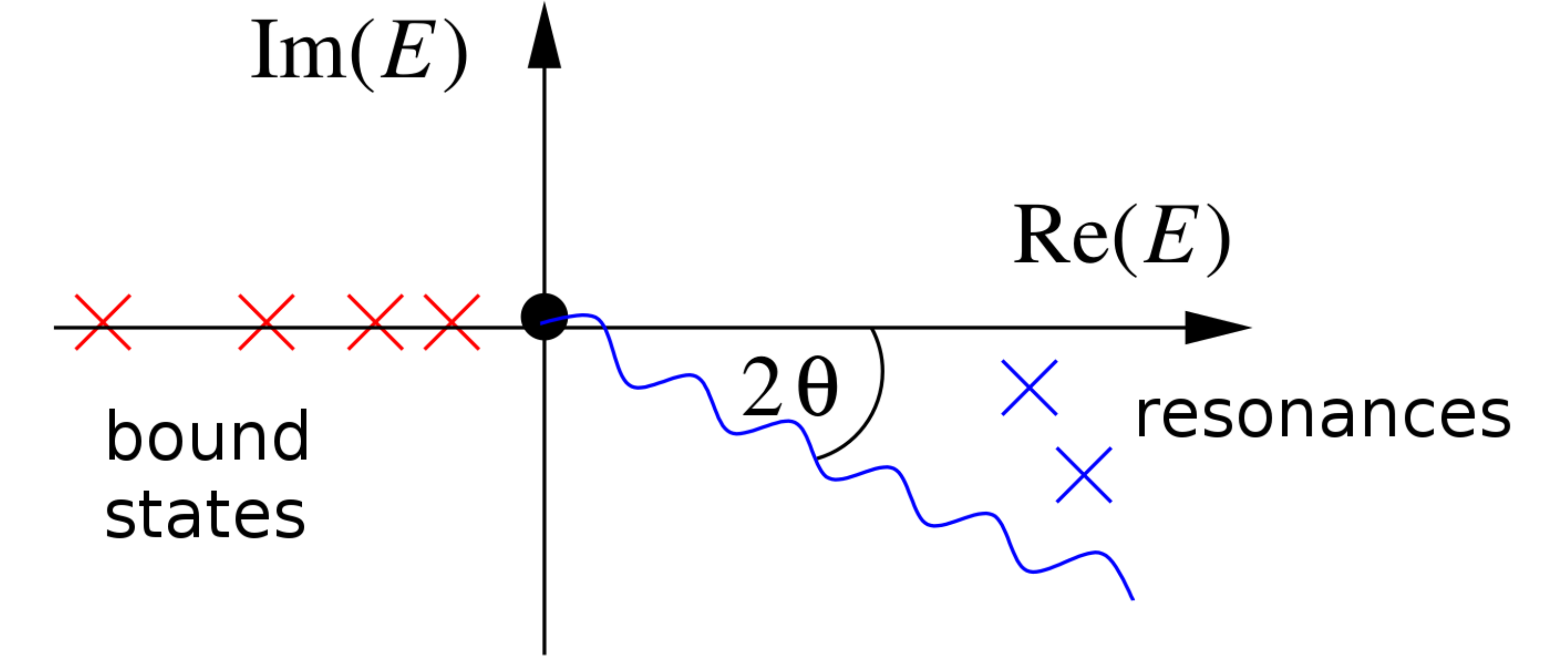}
\caption{Scheme of the complex-coordinate-rotation method.  Resonances
  in the complex energy plane are hidden in Hermitian quantum
  mechanics but can be revealed by the complex-coordinate-rotation
  method.  States representing the continuum are rotated into the
  complex plane by the angle $2\theta$ around the respective threshold.}
\label{fig:ComRot}
\end{figure}

The green exciton states lie above the band gap of the yellow series and are 
coupled to the yellow continuum. Hence, they form quasi-bound resonances rather 
than bound states, even without considering the coupling to the phonons. 
For their description, we introduce complex energies,
whose imaginary part is related to the finite linewidth as
$\gamma =-2\,\mathrm{Im}\,E$.
To compute these eigenenergies, we perform the complex-coordinate rotation
$\boldsymbol{r} \rightarrow \boldsymbol{r}\mathrm{e}^{\mathrm{i}\theta}$ 
\cite{MOISEYEV1998,Reinhardt1982, Ho83}.
It is important to note that, under the complex-coordinate rotation, the 
Hamiltonian \eqref{eq:hamiltonian} becomes a non-Hermitian operator,
and thus allows for complex-valued eigenenergies, as schematically
illustrated in Fig.~\ref{fig:ComRot}.
Continuum states are rotated into the lower complex energy plane,
revealing the resonances, which are \emph{hidden} in a Hermitian
eigenvalue problem.
If the rotation angle $\theta$ is chosen appropriately, the resonance
states become square integrable.
For additional details we refer the reader to 
Refs.~\cite{Zielinski2020,Rommel2020Green}.

\subsection{Central-cell corrections}
\label{sec:central-cell_corrections}
For a correct description of the even-parity exciton states, additional central-cell corrections~\cite{frankevenexcitonseries}
\begin{equation}
  H_{\mathrm{CCC}} = V^{\mathrm{H}} + V_d + H_{\mathrm{exch}}\,,
\label{eq:CCCHaken}
\end{equation}
in the Hamiltonian~\eqref{eq:hamiltonian} are necessary. Here, the Haken potential 

\begin{align}
V^{\mathrm{H}} &=- \frac{\mathrm{e}^2}{4\pi\epsilon_0r}\bigg[ \frac{1}{2\epsilon^\ast_1}\left(e^{-\frac{r}{\rho_{\mathrm{h}1}}} + e^{-\frac{r}{\rho_{\mathrm{e}1}}}\right) \\
& \hspace{10em} + \frac{1}{2\epsilon^\ast_2}\left(e^{-\frac{r}{\rho_{\mathrm{h}2}}} + e^{-\frac{r}{\rho_{\mathrm{e}2}}}\right)\bigg] \nonumber
\end{align}
describes corrections to the
dielectric constant for small exciton radii,
\begin{equation}
V_d = - V_0 V_\mathrm{uc} \delta(\boldsymbol{r})
\end{equation}
is an additional
short distance correction \cite{KavoulakisBohrRadius}, and 
\begin{equation}
H_{\mathrm{exch}}= J_0\left(\frac{1}{4} - \frac{1}{\hbar^2}\boldsymbol{S}_\mathrm{e}\cdot\boldsymbol{S}_\mathrm{h}\right) V_\mathrm{uc} \delta(\boldsymbol{r})
\end{equation}
is the exchange interaction \cite{Uihlein1981}, which causes a splitting of the 
$S$-type states into ortho- and paraexcitons depending on the relative 
orientation of the electron and hole spins.
For the central-cell corrections, we introduce the polaron radii
\begin{equation}
 \rho_{\mathrm{e/h},i} = \sqrt{\frac{\hbar}{2m_\mathrm{e/h}\omega_{\mathrm{LO},i}}} 
\end{equation}
with the energies $\hbar\omega_{\mathrm{LO}i}$ of the longitudinal $\Gamma_4^-$ phonons,
and the values
\begin{equation}
\frac{1}{\epsilon^\ast_i} = \frac{1}{\epsilon_{\mathrm{b}i}} - \frac{1}{\epsilon_{\mathrm{s}i}}\,.
\end{equation}
The parameters $J_0$ and $V_0$ are given in Table~\ref{tab:MaterialParamters}, $V_\mathrm{uc} = a^3$ is
the volume of the unit cell.

Due to the cubic crystal symmetry 
and the associated coupling to angular-momentum states with $\Delta l=\pm 2$, 
these corrections affect not only the $S$ states but also the other even-parity states.
The implementation of these terms requires the calculation of the
complex-rotated matrix elements given in Appendix~D of
Ref.~\cite{frankevenexcitonseries}.
{The difficulty is that these matrix elements form an alternating sum 
of terms with individually very large absolute values.
Thus, an accurate calculation requires the use of a large number of 
significant digits. We therefore work with a computer algebra
system to perform computations to arbitrary precision instead of 
standard double-precision calculations.}
With these preliminaries, we are now in the position to calculate the spectrum 
of even-parity green exciton states.

\subsection{Non-Hermitian generalized eigenvalue problem}
\label{subsec:eigenvalues}
To calculate the eigenstates and eigenvalues of the Hamiltonian 
\eqref{eq:hamiltonian}, we express the time-independent Schr\"odinger equation 
in a complete basis \cite{ImpactValence} using basis states
\begin{equation}
  |\Pi\rangle =
  | N, L, (I, S_\mathrm{h})\, J, F, S_\mathrm{e}, F_\mathrm{t}, M_{F_\mathrm{t}} 
\rangle \,
\label{eq:basis}
\end{equation}
with orbital angular momentum $\boldsymbol{L}$, effective hole spin 
$\boldsymbol{J}$ (as the sum of the quasispin $\boldsymbol{I}$ and the hole spin 
$\boldsymbol{S}_\mathrm{h}$), angular momentum 
$\boldsymbol{F}=\boldsymbol{J}+\boldsymbol{L}$,
and total angular momentum 
$\boldsymbol{F}_\mathrm{t}=\boldsymbol{F}+\boldsymbol{S}_\mathrm{e}$
with its $z$-component $M_{F_\mathrm{t}}$. Here,
$\boldsymbol{S}_\mathrm{e}$ denotes the electron spin.
For the radial part, we use complex rotated Coulomb-Sturmian
functions \cite{CoulombSturmForNuclear},
\begin{equation}
  U_{NL}(r) = N_{NL}(2r/\alpha)^L\mathrm{e}^{-r/\alpha}L_N^{2L+1}(2r/\alpha) \, ,
\label{eq:basis_r}
\end{equation}
which depend on $L$, and are additionally characterized by the radial
quantum number $N$ and the convergence parameter $\alpha$.
The latter can be used for the implementation of the complex scaling
operation, allowing for the calculation of complex resonance states.
To this end, a complex-valued $\alpha = |\alpha|\mathrm{e}^{\mathrm{i}\theta}$
is chosen, resulting in the complex rotation with angle $\theta$.

When expressing the exciton states $|\Psi \rangle$ in the basis~\eqref{eq:basis},
\begin{equation}
|\Psi \rangle = \sum_{\Pi} c_{\Pi} |\Pi \rangle\,,
\label{eq:BasisExpression}
\end{equation}
the Schr\"odinger equation becomes a non-Hermitian generalized eigenvalue 
problem,
\begin{equation}
\boldsymbol{A}\boldsymbol{c} = E \boldsymbol{M} \boldsymbol{c}
\end{equation}
with the Hamiltonian matrix $\boldsymbol{A}_{\Pi'\Pi}=\langle \Pi' | H | \Pi \rangle$,
the overlap matrix $\boldsymbol{M}_{\Pi'\Pi}=\langle \Pi' | \Pi \rangle$ and
the vector $\boldsymbol{c}$ containing the coefficients $c_\Pi$.
Note that the overlap matrix $\boldsymbol{M}$ differs from the identity because 
the Coulomb-Sturmian functions \eqref{eq:basis_r} are not orthogonal.
{To obtain finite matrices and vectors, we introduce cut-offs to the
quantum numbers $N+L+1\leq n_\mathrm{max}$ and $F \leq F_\mathrm{max}$.
These parameters, together with $|\alpha|$ and $\theta$ have to be chosen appropriately to ensure
properly converged results. Good convergence is reached when variations of the parameters do not lead to
significant changes in the calculated spectra.}

{We first diagonalize the Hamiltonian excluding the singular Dirac
delta-terms $V_d$ and $H_\mathrm{exch}$ of the central-cell corrections 
\eqref{eq:CCCHaken}.
From the high-dimensional matrices, we are only interested in a small
window of eigenstates.
For this aim an iterative method is implemented (e.g. in the ARPACK 
package~\cite{arpackuserguide}) that allows for the calculation of eigenvalues
and eigenvectors near a controllable predetermined energy, which is 
numerically more efficient than a direct diagonalization.

After this, we set up a second eigenvalue problem where we include the
delta-terms with only the converged eigenstates from the first
diagonalization.
For this, we diagonalize the entire resulting
low-dimensional eigenvalue problem using a direct LAPACK method~\cite{lapackuserguide3}.}

\section{Dipole transitions between excitonic states}
\label{sec:DipoleTransitions}
In the following, we investigate dipole transitions between different
exciton state which is in contrast to earlier work that focused 
mostly on transitions from the crystal ground state 
\cite{ImpactValence,frankevenexcitonseries,Rommel2020Green}.
The central quantity describing the transition from an initial exciton
state $|\Psi_{\mathrm{i}}\rangle$ to the final exciton state
$|\Psi_{\mathrm{f}}\rangle$ is the transition matrix element
\begin{equation}
  M_{\mathrm{fi}} = 
  \langle \Psi_{\mathrm{f}} | \hat{\boldsymbol{e}}_A \cdot\bpi | \Psi_{\mathrm{i}} \rangle
  \label{eq:fiMatrixelement}
\end{equation}
of the single-photon transition operator, with the polarization direction 
$\hat{\boldsymbol{e}}_A$ of the vector potential associated with the photon 
field
\begin{equation}
\bA(\boldsymbol{x}) = \bA_0 
\mathrm{e}^{\mathrm{i}\boldsymbol{\kappa}\cdot\boldsymbol{x}} \approx \bA_0 = 
A_0 \hat{\boldsymbol{e}}_A
\label{eq:VectorPotential}
\end{equation}
in dipole approximation, where we assume that the momentum
$\hbar\kappa$ of the photon is much smaller than the relative
momentum of exciton and hole.
The operator
\begin{equation}
\bpi = m_0\bv = m_0\frac{\partial\boldsymbol{x}}{\partial t} =  
\frac{\mathrm{i} m_0}{\hbar}\, [\mathcal{H},\boldsymbol{x}]
\end{equation}
denotes the kinetic momentum  operator in a crystal with spin-orbit interaction,
and appears during the minimal-substitution procedure. Note that it differs 
from the quasi-momentum $\boldsymbol{p}$ associated with the Bloch 
eigenfunctions of the band Hamiltonian $\mathcal{H}$. The position operator 
$\boldsymbol{x}$ also has to be distinguished from the coordinates 
$\boldsymbol{r}_\mathrm{e}$ and $\boldsymbol{r}_\mathrm{h}$ that arise from 
the lattice positions in the continuum description of the crystal.

For the interseries transitions discussed in this paper,
$|\Psi_{\mathrm{i}}\rangle$ is mostly a bound yellow exciton state and
$|\Psi_{\mathrm{f}}\rangle$ is an unbound green exciton resonance.
It is therefore sufficient to consider the matrix elements in a basis, e.g., 
with the basis states \eqref{eq:basis}; the transition amplitudes for the
eigenstates can then be obtained by forming appropriate superpositions,
\begin{equation}
  M_{\mathrm{fi}} = \sum_{\Pi',\Pi} c_{\Pi'}^{\mathrm{f}}c_{\Pi}^{\mathrm{i}}
  \langle \Pi' | \hat{\boldsymbol{e}}_A \cdot\bpi | \Pi \rangle\,.
\label{eq:M_fi}
\end{equation}
Note that the coefficients for the left (bra vector) basis states are
not complex conjugated, since they would be real valued without the
complex-coordinate rotation.

\subsection{Operator identity between kinetic momentum and derivatives of the band Hamiltonian}
\label{subsec:Matrixelements} 
We will now derive an operator identity between the kinetic momentum 
$\boldsymbol{\pi}$ and the derivatives of the band Hamiltonian $H(\boldsymbol{p})$
with respect to the momenta.
For that, we consider arbitrary single exciton states, which are effective 
two-particle states of an electron with spin
$\boldsymbol{S}_\mathrm{e,z} = \sigma_{\rm e}$ in the conduction band ($\rmc$) 
and a hole with effective hole spin 
$\boldsymbol{J}_\mathrm{h,z} = \sigma_{\rm h}$ in the valence band ($\rmv$).
An excitonic state with center-of-mass momentum $\boldsymbol{P}$ can then
be written as
\begin{equation}
\left|\Psi_{\tau,\bP}^{\rmc, \rmv}\right\rangle = 
\sum\limits_{\bp}\,\phi_{\tau,\bP} (\bp)\, a^{\dagger}_{\rmc,\sigma_{\rm e}, \bp 
+  \alpha_\mathrm{e}\bP}\, b^{\dagger}_{\rmv,\sigma_{\rm h}, -\bp +  
\alpha_\mathrm{h}\bP}\, \left|\Psi_0\right\rangle \label{eq:state}
\end{equation}
where $\tau=\{N,L,M, \sigma_{\rm e}, \sigma_{\rm h}\}$ is a shorthand notation 
for all the additional quantum numbers of the exciton,
$a^{\dagger}_{\rmc,\sigma,\bq} $ ($b^{\dagger}_{\rmv,\sigma,\bq}$) denotes an 
electron (hole) creation operator and $\left|\Psi_0\right\rangle$ the crystal 
ground state. The coefficients $ \alpha_\mathrm{e}$ and $ \alpha_\mathrm{h}$ 
stem from the transformation from electron-hole coordinates to relative and 
center-of-mass coordinates. They are in principle arbitrary but must
fulfil $ \alpha_\mathrm{e} +  \alpha_\mathrm{h} = 1$. We have chosen the same 
coefficients for all states.

\subsubsection{Dipole approximation}
In the dipole approximation, the excitonic center-of-mass momentum $\bP$  
vanishes and Eq. (\ref{eq:state}) reduces to
\begin{equation}
\left|\Psi_{\tau}^{\rmc, \rmv}\right\rangle = \sum\limits_{\bp}\,\phi_{\tau} 
(\bp)\, a^{\dagger}_{\rmc,\sigma_{\rm e}, \bp }\, b^{\dagger}_{\rmv,\sigma_{\rm 
h}, -\bp}\, \left|\Psi_0\right\rangle .\label{eq:state-dipole} 
\end{equation}
The single-photon transition operator in dipole approximation,
projected onto the Hilbert space spanned by the states in
Eq.~\eqref{eq:state-dipole}, can be written as
\begin{align}
&\frac{e\bA_0\bpi}{m_0} \label{eq:trans-op}\\
&= \frac{e\bA_0}{m_0}\hspace{-.2em}\sum\limits_{\nu, 
\nu'}\,\sum\limits_{\sigma_{\rm e}, \sigma_{\rm e}'}\,\sum\limits_{\bq}\, 
\langle \nu, \sigma_{\rm e}, \bq|\bpi|\nu', \sigma_{\rm e}', \bq\rangle 
\,a^{\dagger}_{\nu,\sigma_{\rm e},\bq}\, a_{\nu',\sigma_{\rm e}',\bq}\nonumber 
\\[5pt]
 &+ \frac{e\bA_0}{m_0}\hspace{-.2em}\sum\limits_{\xi,
   \xi'}\hspace{-.2em}\sum\limits_{\sigma_{\rm h}, \sigma_{\rm
   h}'}\hspace{-.2em}\sum\limits_{\bq}\, \langle \xi, \sigma_{\rm h},
   \bq|\bpi|\xi', \sigma_{\rm h}', \bq\rangle\,
   b^{\dagger}_{\xi,\sigma_{\rm h},\bq}\, b_{\xi',\sigma_{\rm h}',\bq} \nonumber
\end{align}
where $\bA_0$ denotes the vector potential \eqref{eq:VectorPotential}, the 
indices $\nu, \nu'$ sum over all conduction bands, the indices $\xi, \xi'$ sum 
over all valence bands and the $\sigma_{\mathrm{e/h}}$ denote the corresponding 
substates (spins). In the derivation of Eq.~\eqref{eq:trans-op}, the Coulomb 
gauge has been used and the diamagnetic term was ignored{, as it only has
an appreciable influence for very high field strengths
of the incoming electromagnetic wave}. Upon evaluating a 
matrix element of the kind
$e\bA_0\langle\Psi_{\tau}^{\rmc, \rmv}| \bpi|\Psi_{\tau'}^{\rmc', \rmv'}\rangle/m_0$,
there are four cases that must be analyzed separately.

\paragraph*{(i)~Intraseries transitions:} Here, we have $\{\rmc,\sigma_{\rm 
e}\} = \{\rmc',\sigma_{\rm e}'\}$ and $\{\rmv,\sigma_{\rm h}\} = 
\{\rmv',\sigma_{\rm h}'\}$.
Applying the fermionic anti-commutation rules for the creation and annihilation operators, one arrives at
\begin{align}
&\left\langle\Psi_{\tau}^{\rmc, \rmv}\left| \bpi\right|\Psi_{\tau'}^{\rmc, 
\rmv}\right\rangle =\sum\limits_{\bp}\, \phi_{\tau'} 
(\bp)\,\phi_{\tau}^{\dagger} (\bp) \label{eq:dip-trans1}\\
 &\times\,\left(\langle \rmc, \sigma_{\rm e}, \bp|\bpi|\rmc, \sigma_{\rm e}, 
\bp\rangle+ \langle \rmv, \sigma_{\rm h}, -\bp|\bpi|\rmv, \sigma_{\rm h}, 
-\bp\rangle\right). \nonumber 
\end{align}

\paragraph*{(ii)~Hole-driven interseries transitions:} In this case, we have 
$\{\rmc,\sigma_{\rm e}\} = \{\rmc',\sigma_{\rm e}'\}$ but $\{\rmv,\sigma_{\rm 
h}\} \ne \{\rmv',\sigma_{\rm h}'\}$. Hence, we arrive at
\begin{align}
&\left\langle\Psi_{\tau}^{\rmc, \rmv}\left| \bpi\right|\Psi_{\tau'}^{\rmc, 
\rmv'}\right\rangle \label{eq:dip-trans2} \\
&=\sum\limits_{\bp}\, \phi_{\tau'} (\bp)\,\phi_{\tau}^{\dagger} (\bp)\,\langle 
\rmv, \sigma_{\rm h}, -\bp|\bpi|\rmv', \sigma_{\rm h}', -\bp\rangle.\nonumber
\end{align}

\paragraph*{(iii)~Electron-driven interseries transitions:} Here, we have 
$\{\rmc,\sigma_{\rm e}\} \ne \{\rmc',\sigma_{\rm e}'\}$ but $\{\rmv,\sigma_{\rm 
h}\} = \{\rmv',\sigma_{\rm h}'\}$. This case yields
\begin{align}
&\left\langle\Psi_{\tau}^{\rmc, \rmv}\left| \bpi\right|\Psi_{\tau'}^{\rmc', 
\rmv}\right\rangle \label{eq:dip-trans3} \\
 &=\sum\limits_{\bp}\, \phi_{\tau'} (\bp)\,\phi_{\tau}^{\dagger} (\bp)\,\langle 
\rmc, \sigma_{\rm e}, \bp|\bpi|\rmc', \sigma_{\rm e}', \bp\rangle. \nonumber
\end{align}

\paragraph*{(iv)~Two-particle transitions:} In this case, one has 
$\{\rmc,\sigma_{\rm e}\} \ne \{\rmc',\sigma_{\rm e}'\}$ and $\{\rmv,\sigma_{\rm 
h}\} \ne \{\rmv',\sigma_{\rm h}'\}$.
These transitions are forbidden to all orders in single-photon transitions and 
will not be discussed further.

The transitions from the yellow to the green series in {\cuo} are
predominantly hole-driven, although there might be admixture of yellow
states into the green series and vice versa.

\subsubsection{Bloch matrix elements}
The interband matrix elements $\langle n, \sigma, \bp|\bpi|n', \sigma', 
\bp\rangle$, which are expressed in terms of Bloch states, can be written in 
terms of the lattice periodic functions $|u_{n,\sigma},\bp\rangle$ via 
$|n,\sigma, \bp\rangle = e^{\frac{\mathrm{i}}{\hbar}\bp\br}\,|u_{n,\sigma}, 
\bp\rangle$ which results in 
\begin{equation}
\langle n, \sigma, \bp|\bpi|n', \sigma', \bp\rangle = \langle u_{n, \sigma}, 
\bp|\bpi|u_{n', \sigma'}, \bp\rangle + 
\bp\,\delta_{n\,n'}\,\delta_{\sigma\,\sigma'}.
\label{eq:matel-lattice-period}
\end{equation}
Here $n, n'$ denote the bands and $\sigma,\,\sigma'$ the associated spins.
The Hamiltonian 
acting on these lattice periodic functions is the
$\bp\cdot\bpi$-Hamiltonian (usually referred to as the $\bk\cdot\bpi$-Hamiltonian with $\bk = \bp/\hbar$)
\begin{align}
\mathcal{H}_{\bp\cdot\bpi} = \mathcal{H}_0 + \mathcal{H}_{\bp}\,,
\end{align}
with
\begin{align}
\mathcal{H}_0  &= -\frac{\hbar^2\nabla^2}{2m_0}  + V(\boldsymbol{x}) - 
\frac{i\hbar^2}{4m_0^2c^2}\left(\bsi\times\nabla 
V(\boldsymbol{x})\right)\cdot\,\nabla\,, \\
\mathcal{H}_{\bp} &= \frac{\bp}{m_0}\,\cdot\bpi+ \frac{\bp^2}{2m_0}\,,
\end{align}
where $V(\boldsymbol{x})$ is the lattice periodic potential, $\bsi$ the vector 
of Pauli matrices, and $\mathcal{H}_0$ denotes the Hamiltonian at the 
$\Gamma$-point. This implies the relation
\begin{equation}
\bpi  =m_0\bv= m_0 \frac{\partial\mathcal{H}_{\bp\cdot\bpi}}{\partial\bp} - \bp.
\label{eq:bpi-op}
\end{equation}
Inserting Eq. (\ref{eq:bpi-op}) into Eq. (\ref{eq:matel-lattice-period}), we 
arrive at
\begin{equation}
\langle n, \sigma, \bp|\bpi|n', \sigma', \bp\rangle = {m_0}\left\langle u_{n, 
\sigma}, \bp\left| 
\frac{\partial\mathcal{H}_{\bp\cdot\bpi}}{\partial\bp}\right|u_{n', \sigma'}, 
\bp\right\rangle .\label{eq:deriv-ham}
\end{equation}

A perturbation theoretical analysis of Eq.~\eqref{eq:deriv-ham} up to first 
order in $\bp$ has already been performed previously\cite{KruegerInterseries2019},
yielding
{
\begin{equation}
\langle n, \sigma, \bp|\bpi|n', \sigma', \bp\rangle = {m_0}\left\langle u_{n, 
\sigma}, 0\left| \frac{\partial\mathcal{H}_{\bp\cdot\bpi}}{\partial\bp}\right|u_{n', \sigma'}, 
0\right\rangle.
\end{equation}}
{The $\boldsymbol{p}\cdot\boldsymbol{\pi}$ Hamiltonian $\mathcal{H}_{\bp\cdot\bpi}$ describes the
$\boldsymbol{p}$-dependent band dispersion in the crystal. In our system, this is identified with the kinetic energies of the electron
in the conduction band $H_{\rmc}$ and hole in the valence band $H_{\rmv}$ respectively.
Using the kinetic part of the Hamiltonian~\eqref{eq:hamiltonian},
\begin{equation}
T(\bp) = H_{\rmc}(\bp) -H_{\rmv}(\bp) = \frac{\gamma_1'}{2m_{0}}\boldsymbol{p}^2+H_{\rm b}(\boldsymbol{p})\,,
\end{equation}}
{we can summarize} all three cases in Eqs.~\eqref{eq:dip-trans1}-\eqref{eq:dip-trans3}
via
\begin{align}
&\left\langle\Psi_{\tau}^{\rmc, \rmv}\left| \bpi\right|\Psi_{\tau'}^{\rmc, 
\rmv'}\right\rangle \label{eq:combined-transition}\\[10pt]
&={m_0} \sum\limits_{\bp}\,\phi_{\tau}^{\dagger} (\bp)\, \phi_{\tau'} (\bp) 
\left\langle \rmc, \sigma_{\rm e}, \rmv,\sigma_{\rm 
h}\left|\partial_{\bp}T(\bp)\right|\rmc, \sigma_{\rm e}', \rmv',\sigma_{\rm 
h}'\right\rangle\nonumber\\[0pt]
&= m_0\hspace{-.35em}\int\hspace{-.45em}d^3\br\,\psi_{\tau}^{\dagger}(\br)\, 
\hspace{-.2em} \underbrace{\left\langle \rmc, \sigma_{\rm e}, \rmv,\sigma_{\rm 
h}\left|\partial_{\bp}T(\bp)\right|\rmc, \sigma_{\rm e}', \rmv',\sigma_{\rm 
h}'\right\rangle}_{\mathcal{O}(\bp)}\hspace{-.2em}\,\psi_{\tau'}(\br),\nonumber
\end{align}
where {the matrix element is evaluated} in the twelve-dimensional basis of 
electron-hole spin-states $|\rmc,\sigma_{\rm e}, \rmv, \sigma_{\rm h}\rangle$.
The second line gives the equivalent expression in real space, where 
$\psi_{\tau}(\br)$ is the real-space envelope function of the state 
$|\Psi_{\tau'}^{\rmc, \rmv'}\rangle$.
These states span the same Hilbert space as the basis states \eqref{eq:basis}.
Noting that only the kinetic energy terms in the Hamiltonian 
\eqref{eq:hamiltonian} contain the relative momentum operator $\boldsymbol{p}$, 
we obtain the identity
\begin{equation}
 \boldsymbol{\pi} = m_0 \frac{\partial}{\partial \boldsymbol{p}} 
H(\boldsymbol{p})
 \label{eq:MatrixlementPiOperator}
\end{equation}
valid for the one-exciton states considered in this paper.
Equation~\eqref{eq:MatrixlementPiOperator} is an operator identity in
the one-exciton Hilbert space spanned, e.g., by the basis \eqref{eq:basis},
and is valid {for vanishing center-of-mass momentum $\boldsymbol{P}$ and relative momentum
$\boldsymbol{p}$ much smaller than the extent of the Brillouin zone.}

\subsection{Numerical evaluation of the matrix elements $\langle \Pi' |\pi_z| 
\Pi\rangle$}
\label{subsec:NumericalEvaluation}
The computation of the dipole transition matrix elements
$M_{\mathrm{fi}}$ in Eq.~\eqref{eq:M_fi} requires one to evaluate the
matrix elements $\langle \Pi' | \bpi | \Pi\rangle$ of the
operator \eqref{eq:MatrixlementPiOperator} in the
basis \eqref{eq:basis}.

From Eq.~\eqref{eq:MatrixlementPiOperator} we obtain
\begin{align}
\frac{\boldsymbol{\pi}}{m_0} =
\frac{\partial}{\partial \boldsymbol{p}}H(\boldsymbol{p}) 
= \frac{\gamma_1'\boldsymbol{p}}{m_0} + \frac{\partial 
H_\mathrm{b}(\boldsymbol{p})}{{\partial \boldsymbol{p}}}\,.
\label{eq:DerivativeHamiltonian}
\end{align}
We focus on the component $\pi_z$ for light polarized along the $z$-axis.
The matrix elements for $p_z$ in the basis \eqref{eq:basis} are derived in 
Appendix~\ref{sec:MatrixElementPz}.
The more difficult part is to evaluate the second term in
Eq.~\eqref{eq:DerivativeHamiltonian}.
Instead of deriving the expression in detail here,
we connect this problem to terms already calculated 
in Ref.~\cite{frankjanpolariton}.
They consider the Hamiltonian \eqref{eq:hamiltonian} in center-of-mass 
coordinates with a nonvanishing center-of-mass momentum $P = \hbar K$ parallel 
to a given axis. Here, we are interested in the case $\boldsymbol{P} \parallel 
[001]$ related to the derivative with respect to $p_z$. 
This means that we can set $\boldsymbol{P} = P\boldsymbol{e}_z$ in the 
following.
Following Ref.~\cite{frankjanpolariton}, we expand the Hamiltonian in powers of 
$P$ as
\begin{equation}
H(\boldsymbol{p},\boldsymbol{P}) = H_0 + P\,H_1 + P^2\,H_2\,.
\end{equation}
The center-of-mass transformation \eqref{eq:COMcoordinates} is chosen in such a 
way that terms linear in $\boldsymbol{P}$ vanish without the corrections from 
the valence band. This means that the term $H_1$ arises solely from the kinetic 
energy $H_\mathrm{h}$ of the hole. More explicitly, we can write
\begin{align}
H_\mathrm{h}&(\boldsymbol{p}_\mathrm{h} = -\boldsymbol{p} + 
\alpha_\mathrm{h}\boldsymbol{P})\\
 &= \frac{\boldsymbol{p}^2}{2m_\mathrm{h}} + H_\mathrm{b}(\boldsymbol{p}) 
-\frac{\alpha_\mathrm{h}P}{m_\mathrm{h}}p_z + P\,H_1 + \mathcal{O}(P^2)\,, 
\nonumber
\end{align}
where $\alpha_h = m_\mathrm{h}/(m_\mathrm{h}+m_\mathrm{e})$ is determined by 
the center-of-mass transformation. We first differentiate both sides with 
respect to $P = P_z$ and evaluate at $P = 0$,
\begin{equation}
\alpha_\mathrm{h} \frac{\partial H_\mathrm{h}}{\partial {p}_{\mathrm{h},z}} 
(\boldsymbol{p}_\mathrm{h} = -\boldsymbol{p})
 = -\alpha_\mathrm{h} \frac{\partial H_\mathrm{h}}{\partial {p}_z} 
(-\boldsymbol{p})
 = -\frac{\alpha_\mathrm{h}}{m_\mathrm{h}}p_z + H_1\,.
\end{equation}
On the other hand, first setting $P = 0$ and differentiating with respect to 
$p_z$ leads to
\begin{equation}
\frac{\partial H_\mathrm{h}}{\partial {p}_z} (-\boldsymbol{p})
 = \frac{p_z}{m_\mathrm{h}} + \frac{\partial H_\mathrm{b}(\boldsymbol{p})}{{\partial p_z}}\,.
\end{equation}
Comparing these results, we obtain the identity
\begin{equation}
\frac{\partial H_\mathrm{b}}{{\partial p_z}}(\boldsymbol{p}) = -\frac{1}{\alpha_\mathrm{h}} H_1 = -\frac{m_\mathrm{e} \gamma_1'}{m_0} H_1\,.
\end{equation}
Inserted into Eq.~\eqref{eq:DerivativeHamiltonian}, we finally find
\begin{equation}
 \pi_z = \gamma_1'\left( p_z - m_\mathrm{e}H_1 \right)\,,
\end{equation}
with~\cite{frankjanpolariton}
\begin{align}
&H_{1}= -\frac{1}{2\hbar^{2}m_{\mathrm{e}}}\bigg\{ 2\sqrt{\frac{5}{3}}\mu'\left[P^{(1)}\times I^{(2)}\right]_{0}^{(1)} \\
&\hspace{12em}+4\sqrt{\frac{2}{5}}\delta'\left[P^{(1)}\times I^{(2)}\right]_{0}^{(3)}\bigg\} \nonumber \\
&\hspace{-.2em}-\hspace{-.2em}\frac{3\eta_{1}}{\gamma{}_{1}'\hbar^{2}m_{\mathrm{e}}}\bigg\{ \frac{2}{3}P_{0}^{(1)}\left(I^{(1)}\cdot S_{\mathrm{h}}^{(1)}\right)+2\sqrt{\frac{5}{3}}\nu\left[P^{(1)}\times D^{(2)}\right]_{0}^{(1)}\nonumber \\
&\hspace{12em}+4\sqrt{\frac{2}{5}}\tau\left[P^{(1)}\times D^{(2)}\right]_{0}^{(3)}\bigg\}, \nonumber 
\end{align}
using the abbreviations
\begin{equation}
D_{k}^{(2)}=\left[I^{(1)}\times S_{\mathrm{h}}^{(1)}\right]_{k}^{(2)}
\end{equation}
and 
\begin{align}
\mu'=\frac{6\gamma_{3}+4\gamma_{2}}{5\gamma'_{1}},\quad\delta'=\frac{\gamma_{3}-\gamma_{2}}{\gamma'_{1}},\nonumber\\
\nu=\frac{6\eta_{3}+4\eta_{2}}{5\eta_{1}},\quad\tau=\frac{\eta_{3}-\eta_{2}}{\eta_{1}}\,.
\end{align}
All relevant matrix elements can be found in Ref.~\cite{frankjanpolariton}.

\section{Results and discussion}
\label{sec:Results}
\begin{figure*}[t]
\includegraphics[width=\columnwidth]{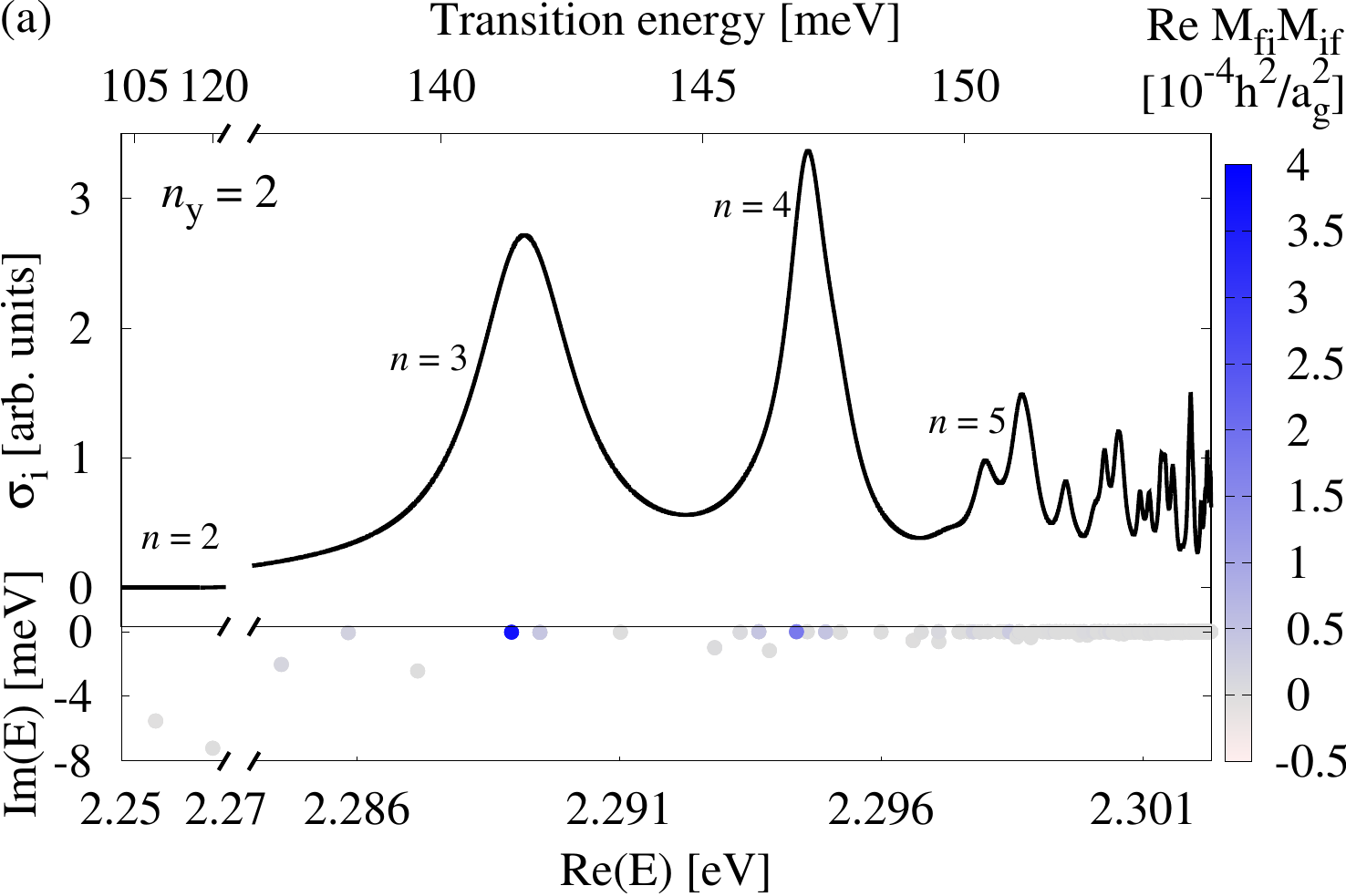}
\includegraphics[width=\columnwidth]{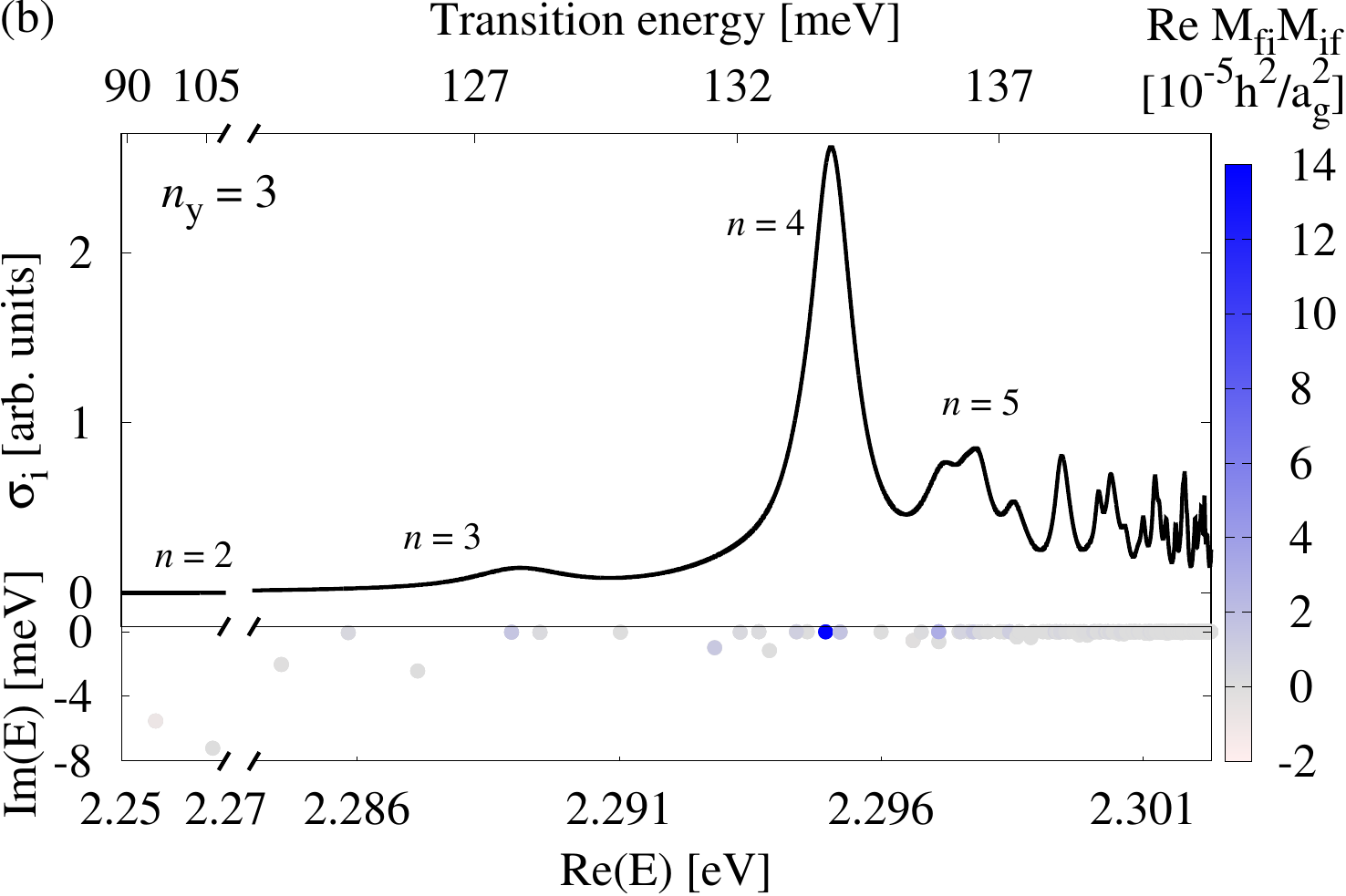}\\
\includegraphics[width=\columnwidth]{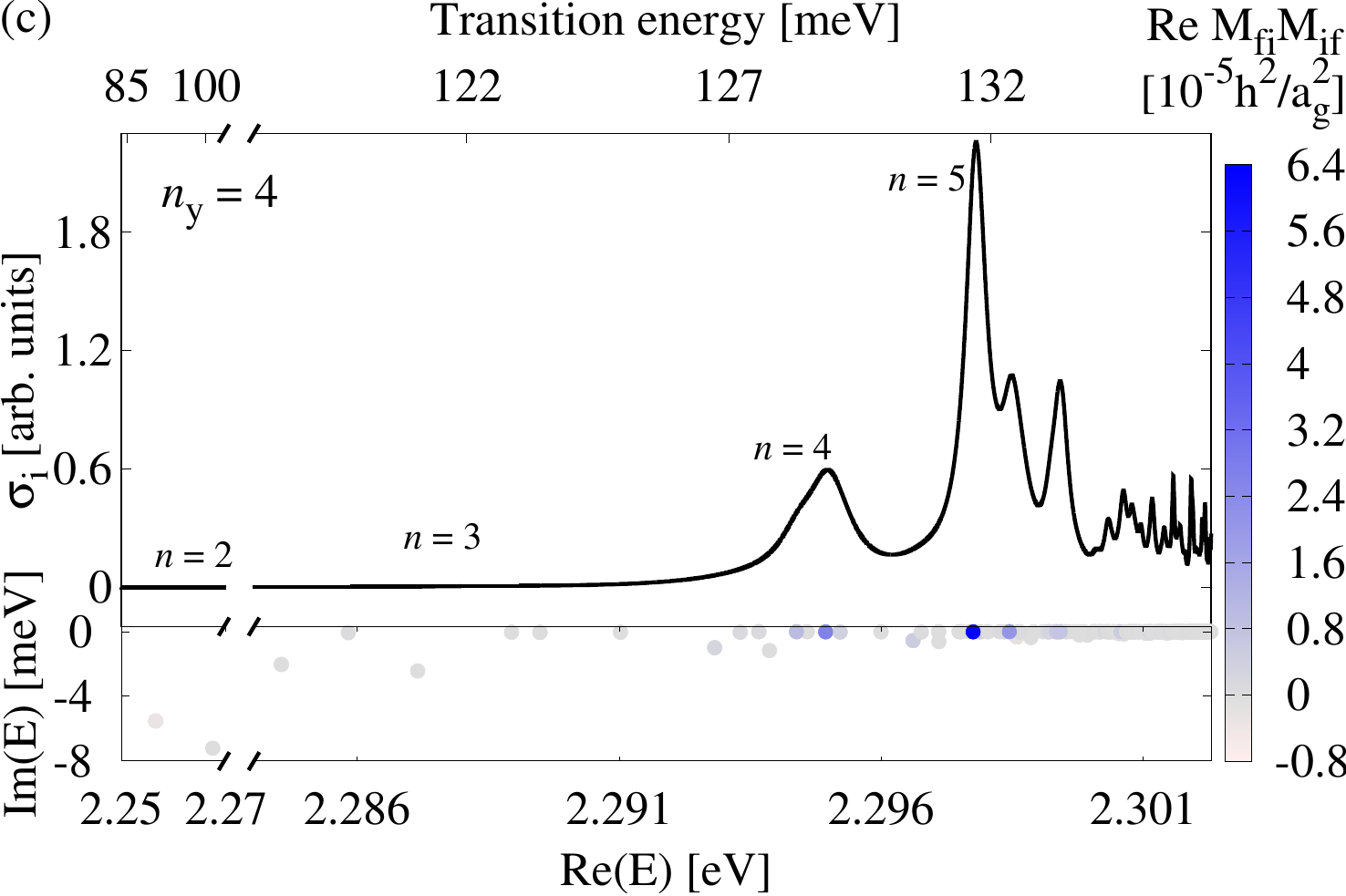}
\includegraphics[width=\columnwidth]{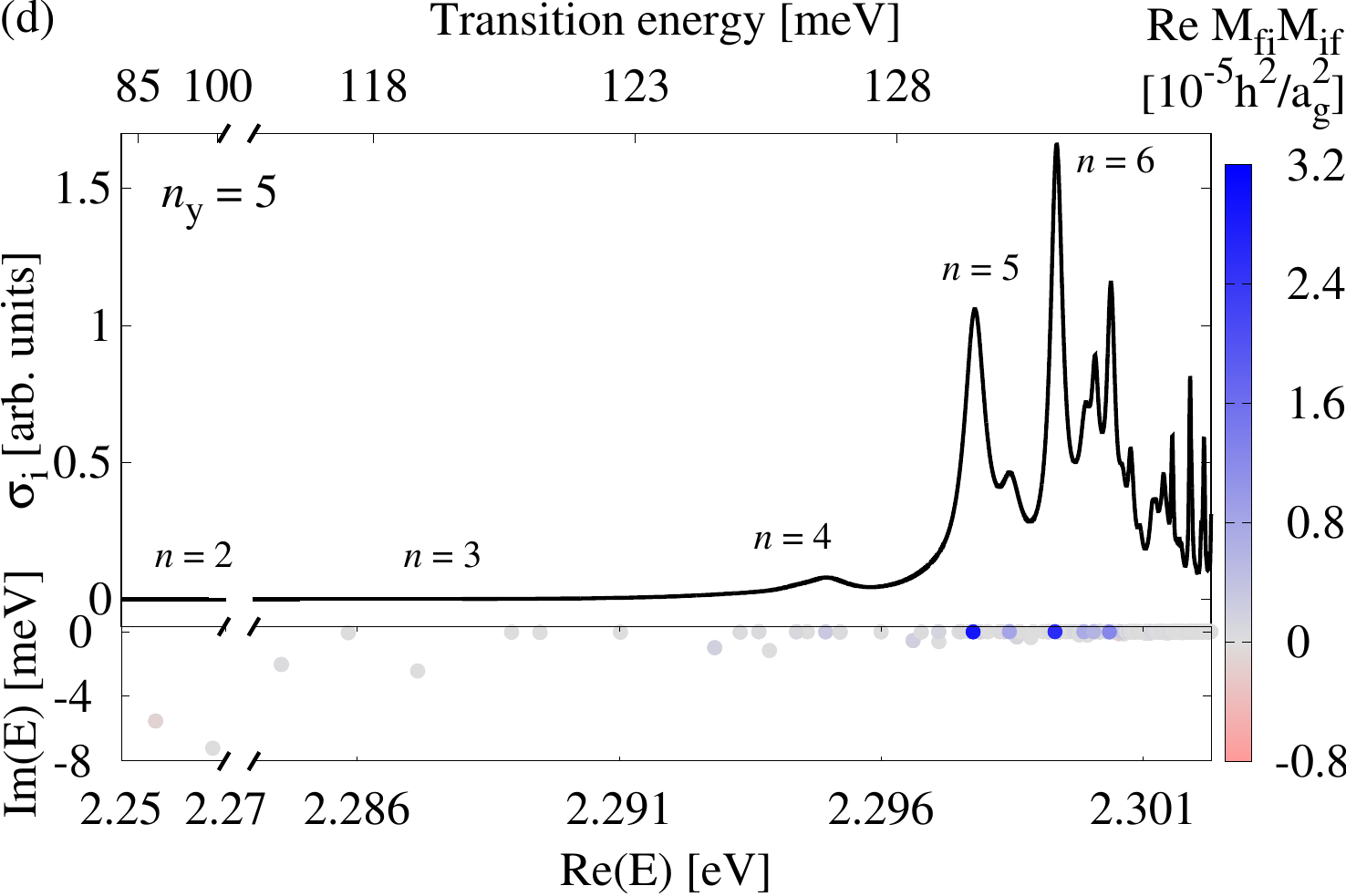}
\caption{Spectrum of transitions between odd parity yellow $P$ excitons transforming like the function $z$ and
even parity green states. The transition is mediated by photons polarized in the $z$-direction. In the top part of the panels,
we show the spectrum using the linewidths derived from the complex rotation
corrected by Eq.~\eqref{eq:phononlinewidth}
to incorporate the influence of the phonons. The uncorrected complex energy is presented in the bottom part
of the panel. The color additionally
shows the real part of the square $M_\mathrm{fi}M_\mathrm{if}$ of the interseries transitions matrix element
introduced in Eq.~\eqref{eq:fiMatrixelement},
which is proportional to the complex generalization of the oscillator strength.}
\label{fig:OddToEvenZZ}
\end{figure*}
\begin{table}[]
\caption{Real (R) and imaginary (I) parts of squared transition matrix elements 
$M^2 = M_\mathrm{if}M_\mathrm{fi}$ in units of 
$10^{-6}\,\mathrm{h}^2\,\mathrm{a}_\mathrm{g}^{-2}$ for certain selected green 
exciton states of Energy $E$ = Re\,$E_\mathrm{f}$. The initial odd parity 
yellow $P$ state of irreducible representation $\Gamma_4^-$ transforms like $z$ 
and the light is polarized along the $z$-direction.}
\begin{tabular}{r|rr|rr|rr|rr}
 & $2P$ &  & $3P$ &  & $4P$ &  & $5P$ & \\[-1em]
\ac{$E$\,[eV]} & R\,$M^2$ & I\,$M^2$ & R\,$M^2$ & I\,$M^2$ & R\,$M^2$ & I\,$M^2$ & R\,$M^2$ & I\,$M^2$ \\
\hline
2.28456 & 15.02 & -19.56 & -1.36 & 1.11 & 0.11 & -1.55 & 0.32 & -0.71\\[-1em]	
2.28583 & 22.59 & -0.60 & 3.92 & -0.12 & 0.66 & -0.03 & 0.23 & -0.01\\[-1em]
2.28895 & 366.60 & 1.17 & 15.29 & 0.41 & 0.17 & 0.03 & 	0.00 & 0.00\\[-1em]
2.28949 & 41.06 & 1.24 & 	2.22 & 0.10 & 0.03 & 0.01 & 0.00 & 0.00\\[-1em]
2.29283 & 3.00 & 7.36 & 13.42 & -10.83 & 2.20 & -2.18 & 1.87 & -0.88\\[-1em]
2.29367 & 43.45 & -0.67 & 1.61 & -0.11 & 1.25 & -0.05 & 0.47 & -0.02\\[-1em]
2.29439 & 180.87 & 1.34 & 8.95 & 1.14 & 9.62 & 0.29 & 0.85 & 0.06\\[-1em]
2.29494 & 46.04 & -0.51 & 139.46 & 0.42 & 26.58 & 0.50 & 3.05 & 0.15\\[-1em]
2.29522 & 0.82 & 0.08 & 16.55 & 0.80 & 3.92 & 0.09 & 0.34 & 0.03\\[-1em]
2.29710 & 9.91 & -0.04 & 30.51 & -0.08 & 0.03 & 0.02 & 1.32 & 0.06\\[-1em]
2.29776 & 19.31 & -0.29 & 12.41 & -0.22 & 62.81 & -0.05 & 29.01 & 0.28\\[-1em]
2.29845 & 33.28 & 0.18 & 9.36 & 0.55 & 21.16 & -0.69 & 8.31 & 0.52\\[-1em]
2.29864 & 16.24 & 0.73 & 1.87 & 0.02 & 4.49 & -0.43 & 0.20 & 0.10\\[-1em]
2.29932 & 10.08 & -0.18 & 6.92 & -0.11 & 2.06 & -0.08 & 25.78 & -0.20\\[-1em]
\end{tabular}
\label{tab:OddToEvenZZ}
\end{table}
In the following, we present our results for the dipole transition 
probabilities for two cases of interseries transitions. As parity is an exact 
quantum number, we separately discuss transitions first from odd parity to even 
parity states, and then from even parity to odd parity states. We choose a 
coordinate system where the $x$-, $y$- and $z$-axes are parallel to $[100]$, 
$[010]$, and $[001]$ directions, respectively.

\subsection{Interseries absorption spectra}
\label{subsec:photocrosssection}
The transition matrix elements $M_{\mathrm{fi}}$ can be used to
calculate interseries absorption spectra.
The photoabsorption cross section $\sigma_{\mathrm{i}}$ from the
initial state $|\Psi_{\mathrm{i}}\rangle$ at the spectral position
$E=\hbar\omega_\mathrm{ph}$ is given by \cite{Zielinski2020,Rescigno1975}
\begin{equation}
\sigma_{\mathrm{i}}(\omega_{\mathrm{ph}}) = \frac{4\pi\alpha\hbar}{m_0^2 
\omega_\mathrm{ph}}\mathrm{Im}
\sum_{\mathrm{f}} 
\frac{M_{\mathrm{fi}}M_{\mathrm{if}}}{E_{\mathrm{f}}-E_{\mathrm{i}}
-\hbar\omega_\mathrm{ph}}\,,
\label{eq:Photocrosssection}
\end{equation}
with the fine-structure constant $\alpha$ and
$\hbar\omega_\mathrm{ph} \approx E_{\mathrm{f}}-E_{\mathrm{i}}$.
Note that in general, $M_{\mathrm{fi}} \ne M_{\mathrm{if}}^\ast$ for complex rotated states, and thus the numerator
in Eq.~\eqref{eq:Photocrosssection} does not simplify to $|M_{\mathrm{fi}}|^2$.

To avoid extremely narrow peaks for certain states, we phenomenologically
model an additional linewidth caused by the coupling to phonons in the crystal.
In a simplified model, the phonon-induced linewidth has a power-law dependency 
on the principal quantum number $n$ as \cite{GiantRydbergExcitons,Toyozawa1958}
\begin{align}
\gamma_\mathrm{ph}(n) = \gamma^\mathrm{ph}_0 n^{-3}\,.
\label{eq:phononlinewidth}
\end{align}
We estimate the parameter $\gamma^\mathrm{ph}_0 = 56.4\,\mathrm{meV}$ and 
assign to each resonance an effective quantum number $n_\mathrm{eff}$ based on 
the real part of its energy as outlined in Appendix~\ref{sec:PhononLinewidths}.
The resulting linewidth shifts the imaginary part of the complex energy according to
$E_\mathrm{f}$ $\rightarrow$ $E_\mathrm{f} - \mathrm{i} \gamma_\mathrm{ph} /2$.

\subsection{Transitions from odd-parity yellow exciton states to even-parity 
green states}
\begin{figure*}
\includegraphics[width=\columnwidth]{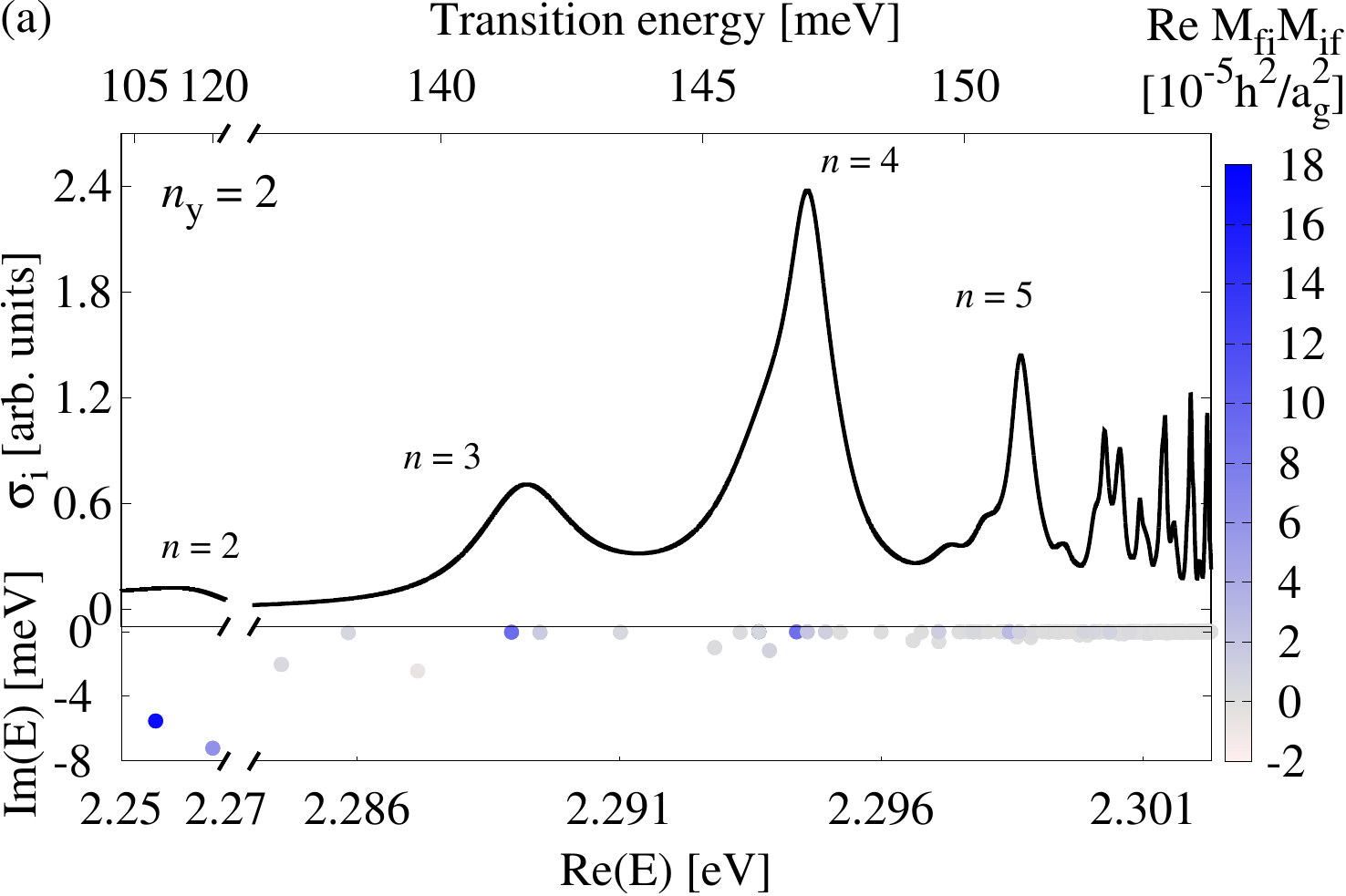}
\includegraphics[width=\columnwidth]{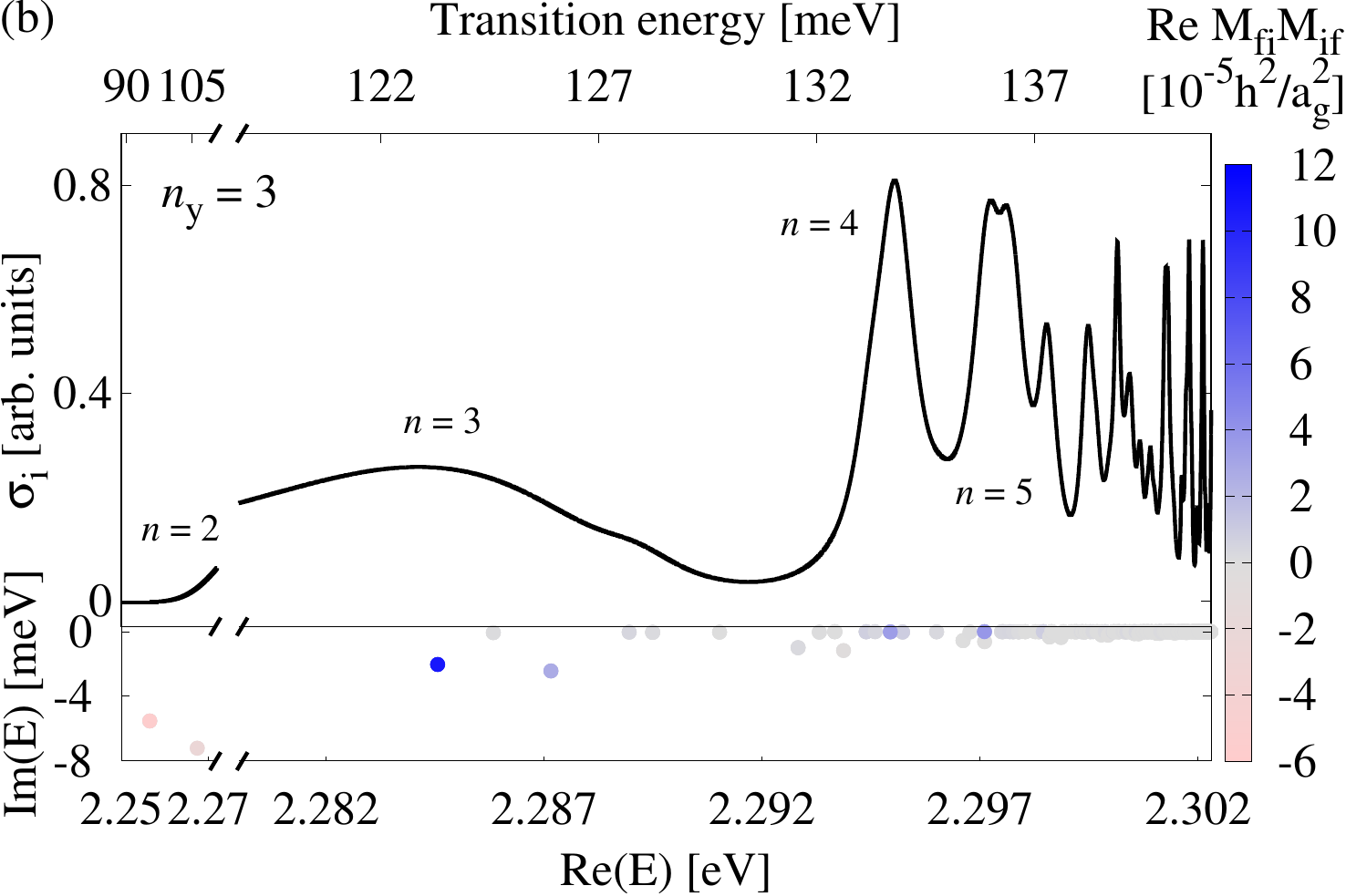}
\includegraphics[width=\columnwidth]{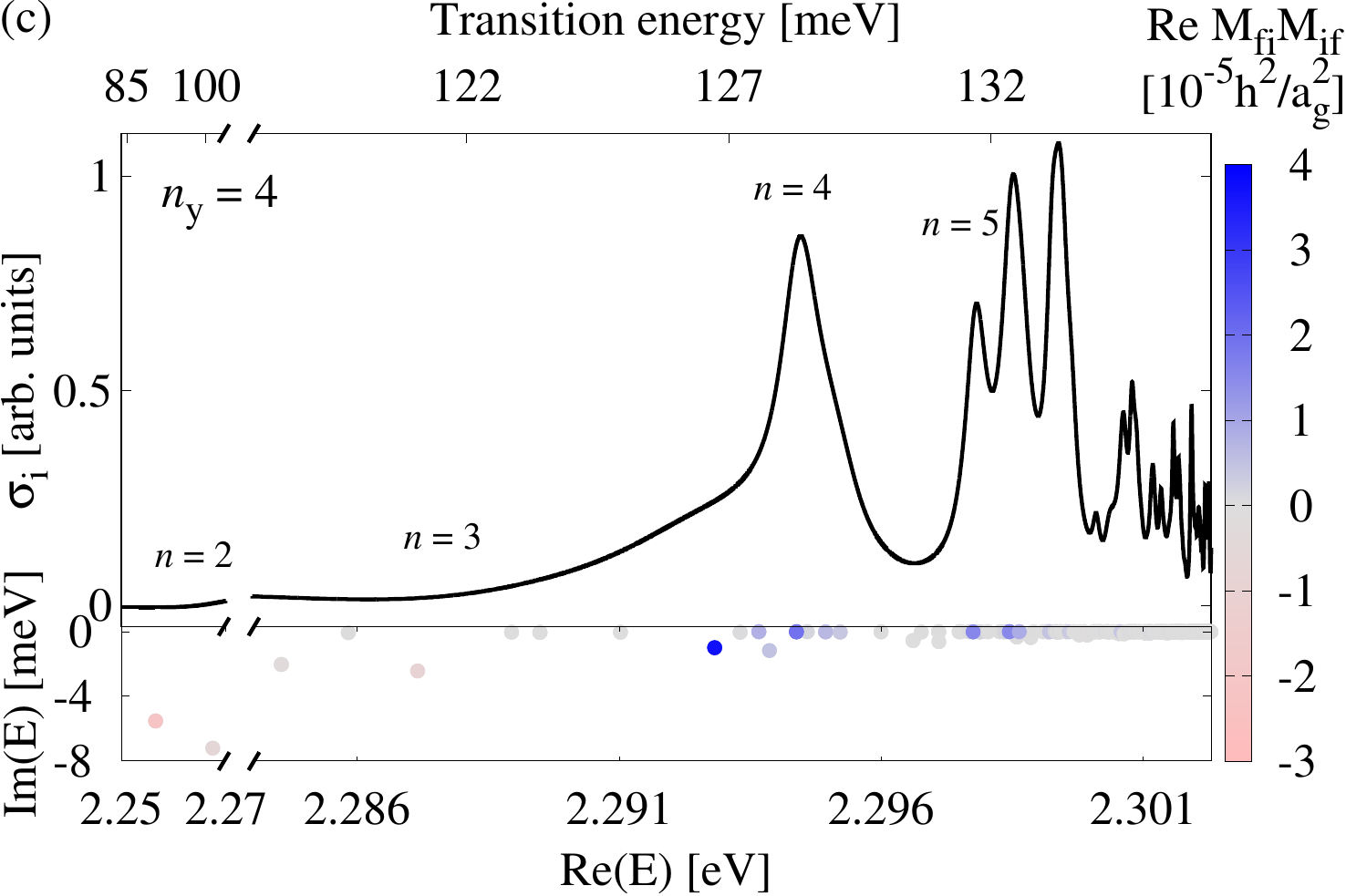}
\includegraphics[width=\columnwidth]{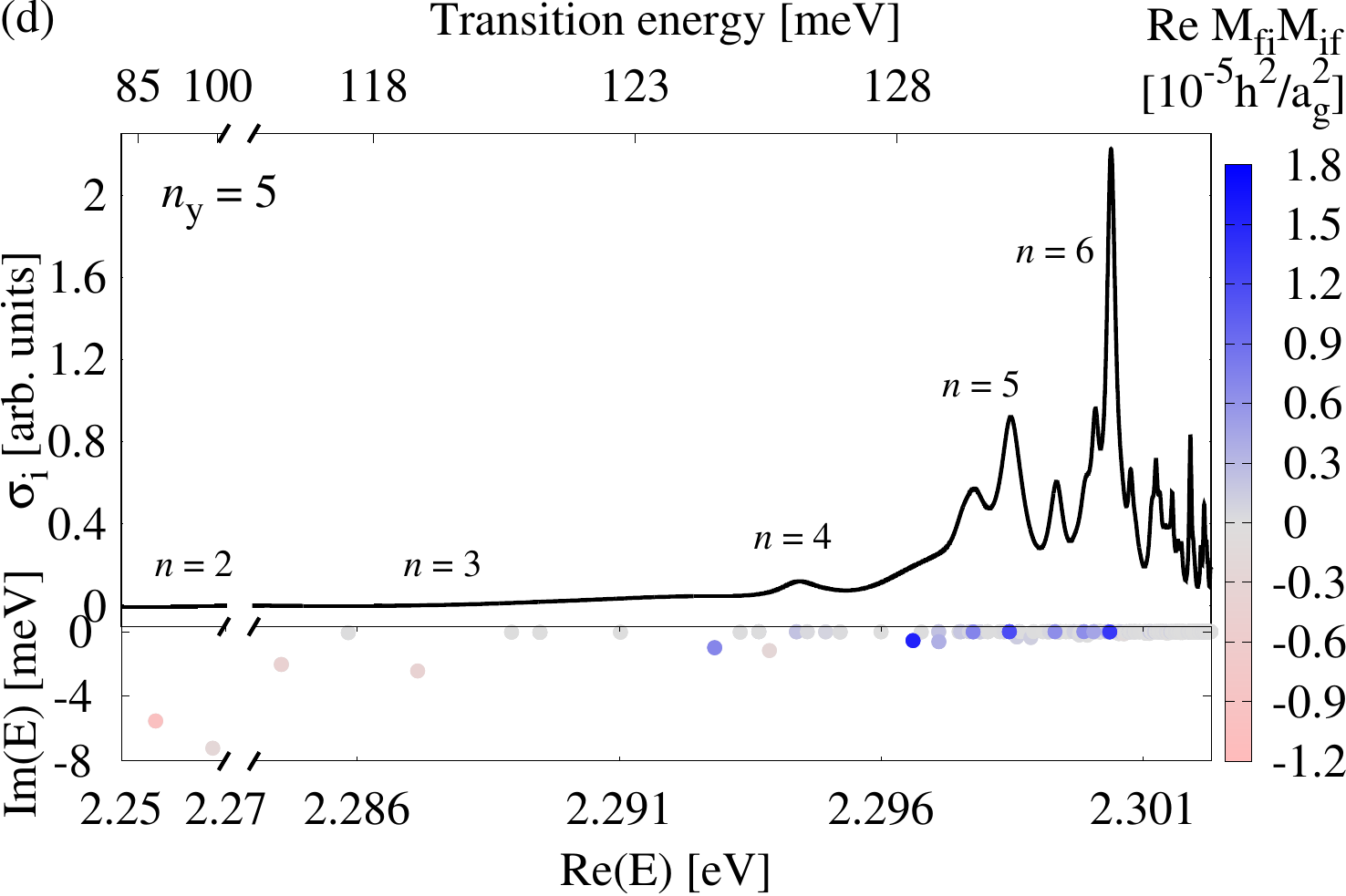}
\caption{Same as Fig.~\ref{fig:OddToEvenZZ}, but the initial odd parity yellow $P$ states transform like the function $y$.}
\label{fig:OddToEvenYZ}
\end{figure*}

\begin{table*}[]
\caption{Same as Table~\ref{tab:OddToEvenZZ} but the initial odd parity
yellow $P$ states transform like $y$.}
\begin{tabular}{r|rr|rr|rr|rr}
 & $2P$ &  & $3P$ &  & $4P$ &  & $5P$ & \\[-1em]
\ac{$E$\,[eV]} & R\,$M^2$ & I\,$M^2$ & R\,$M^2$ & I\,$M^2$ & R\,$M^2$ & I\,$M^2$ & R\,$M^2$ & I\,$M^2$ \\
\hline \\[-7ex]
2.25655 & 170.05 & 295.75 & -55.84 & -10.02 & -22.11 & -16.43 & -10.17 & -10.68\\[-1.2em]
2.26745 & 61.88 & 149.44 & -24.63 & -32.42 & -5.93 & -15.23 & -2.25 & -7.71\\[-1.2em]
2.28456 & 4.52 & 13.21 & 107.45 & 70.38 & -2.65 & 18.35 & -4.21 & 5.47\\[-1.2em]
2.28716 & -7.42 & 6.12 & 27.30 & 44.18 & -9.08 & 1.49 & -3.92 & -1.10\\[-1.2em]
2.28895 & 91.65 & 0.29 & 3.82 & 0.10 & 0.04 & 0.01 & 0.00 & 0.00\\[-1.2em]
2.29283 & 1.98 & 2.69 & 2.46 & 1.01 & 37.11 & 14.39 & 7.05 & 9.05\\[-1.2em]
2.29367 & 116.25 & 0.95 & 1.11 & -0.06 & 8.16 & -0.21 & 1.67 & -0.03\\[-1.2em]
2.29439 & 9.83 & -0.06 & 2.46 & -0.17 & 18.61 & -1.63 & 2.42 & -0.03\\[-1.2em]
2.29439 & 89.72 & 3.11 & 7.48 & 0.58 & 19.32 & -1.38 & 2.23 & 0.04\\[-1.2em]
2.29494 & 11.51 & -0.13 & 34.86 & 0.11 & 6.64 & 0.12 & 0.76 & 0.04\\[-1.2em]
2.29661 & -0.59 & 0.36 & -0.07 & 0.23 & -0.27 & -0.12 & 15.41 & -2.43\\[-1.2em]
2.29710 & 13.46 & 0.16 & 38.08 & 0.20 & 0.24 & 0.00 & 2.30 & 0.21\\[-1.2em]
2.29776 & 4.83 & -0.07 & 3.10 & -0.06 & 15.70 & -0.01 & 7.25 & 0.07\\[-1.2em]
2.29845 & 11.70 & 0.75 & 3.82 & 0.27 & 4.92 & -0.06 & 9.73 & 0.70\\[-1.2em]
2.29845 & 28.58 & 1.00 & 8.52 & 0.61 & 15.28 & -0.49 & 11.81 & 0.46\\[-1.2em]
2.30037 & 8.11 & 0.23 & 2.82 & 0.29 & 0.19 & -0.13 & 13.45 & -1.78\\[-1.2em]
\end{tabular}
\label{tab:OddToEvenYZ}
\end{table*}

\begin{figure*}
\includegraphics[width=\columnwidth]{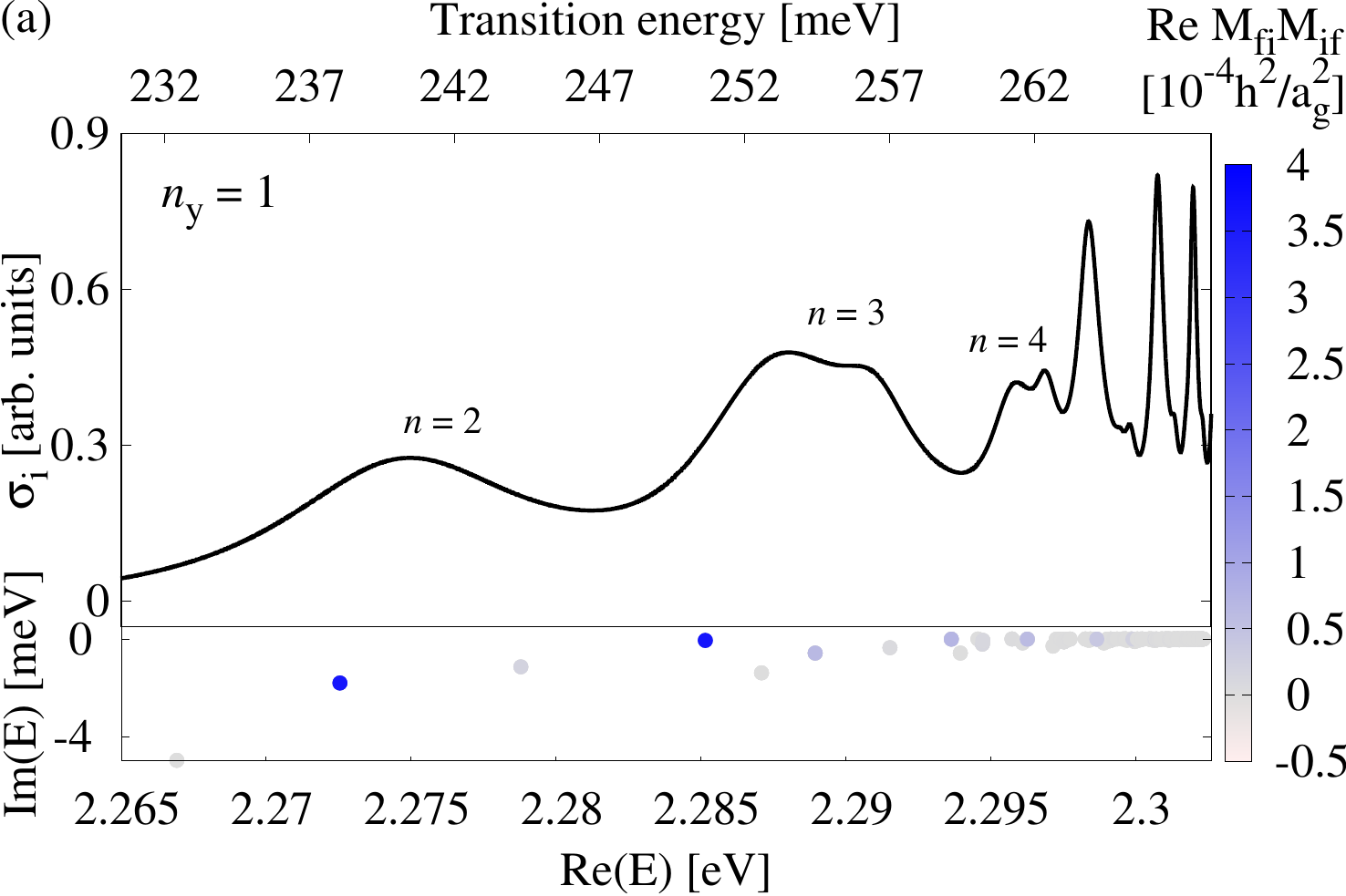}
\includegraphics[width=\columnwidth]{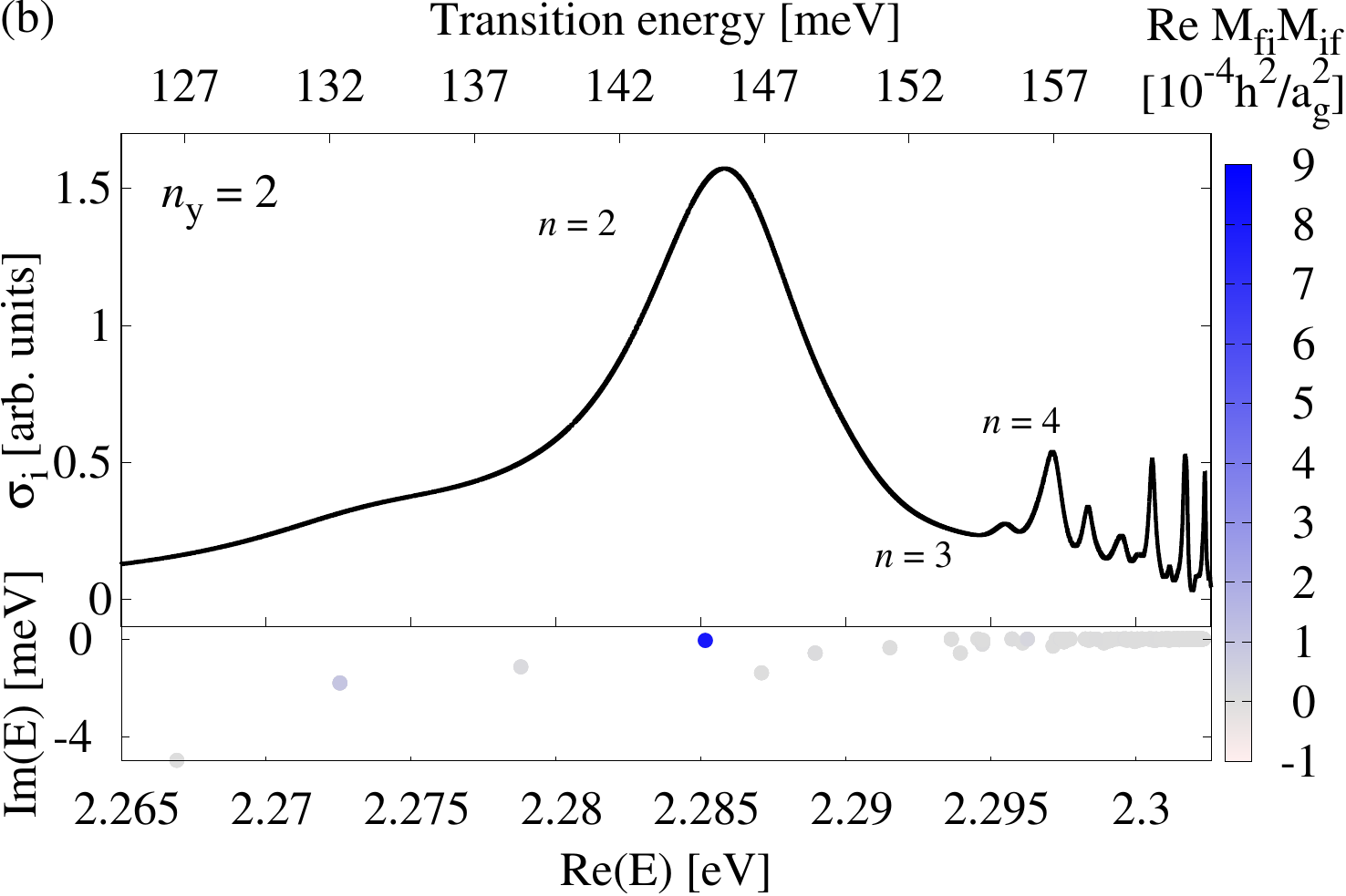}
\includegraphics[width=\columnwidth]{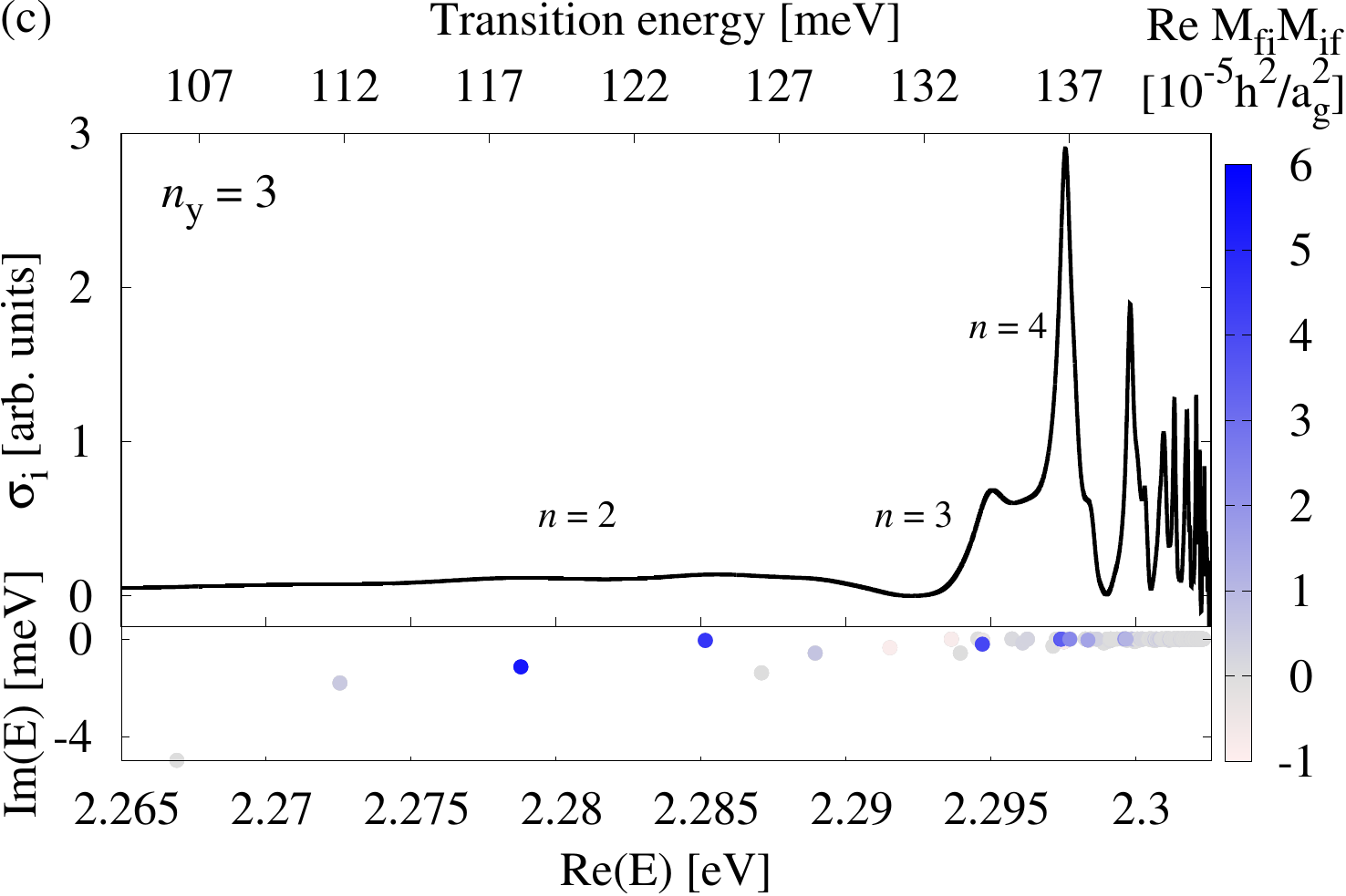}
\includegraphics[width=\columnwidth]{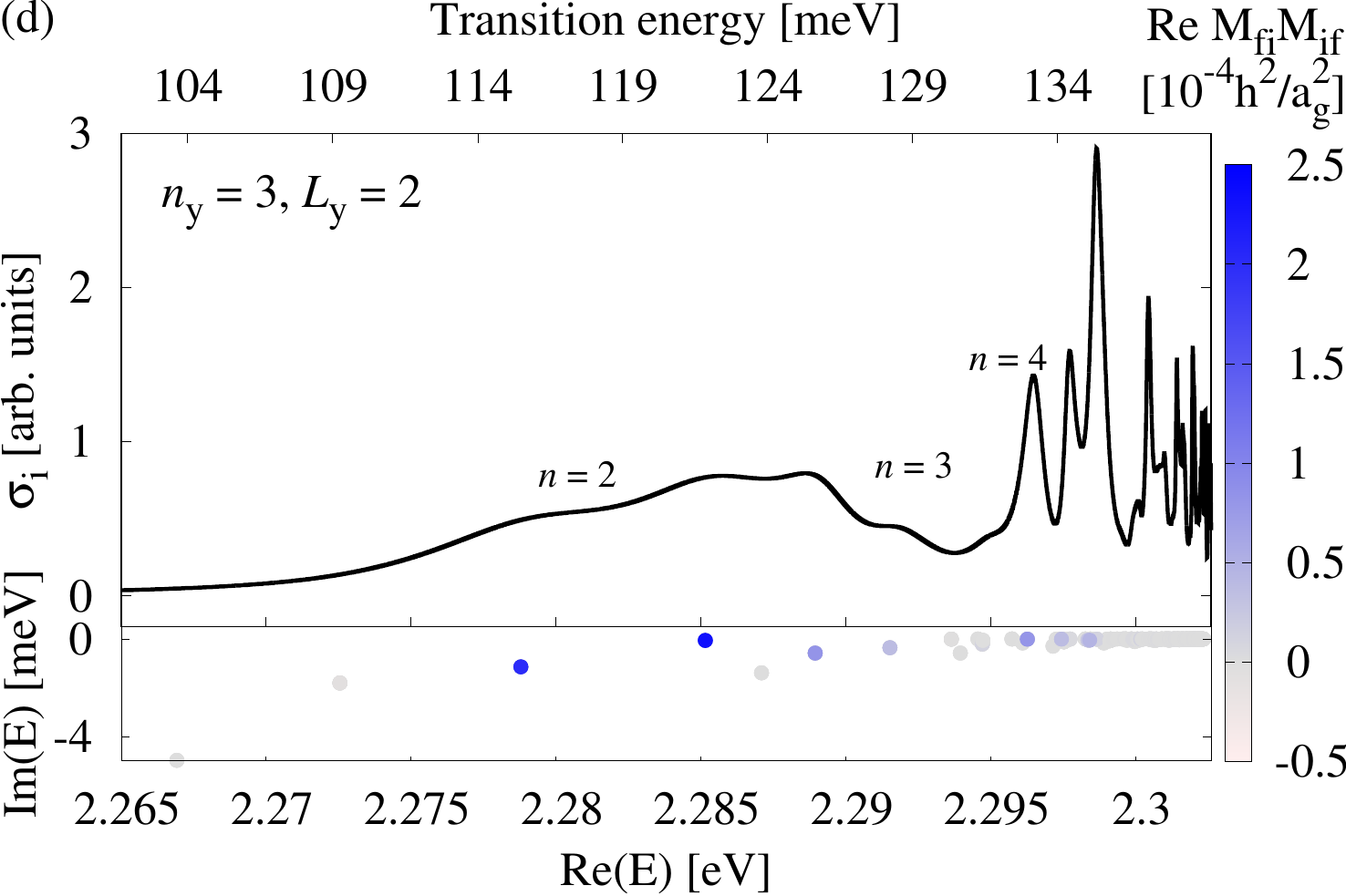}
\caption{Same as Fig.~\ref{fig:OddToEvenZZ}, but the initial even parity states are S ((a)-(c)) and D (d) states transforming like the function $xy$
of the irreducible representation $\Gamma_5^+$.}

\label{fig:EvenToOdd}
\end{figure*}

\begin{table*}
\caption{Same as Table~\ref{tab:OddToEvenZZ} but the initial even parity
yellow states transforms like $xy$ of the irreducible representation $\Gamma_5^+$.
In the last two columns we show the transition matrix elements for the initial \emph{green} $1S$ exciton state.}
\begin{tabular}{r|rr|rr|rr|rr|rr}
 & $1S$ &  & $2S$ &  & $3S$ &  & $3D$ &  & $1S$g & \\[-1em]
\ac{$E$\,[eV]} & R\,$M^2$ & I\,$M^2$ & R\,$M^2$ & I\,$M^2$ & R\,$M^2$ & I\,$M^2$ & R\,$M^2$ & I\,$M^2$ & R\,$M^2$ & I\,$M^2$ \\
\hline \\[-7ex]
2.27254	& 359.83 & 	-18.09	& 96.55	& -2.35	& 5.32	& 25.33	& -6.76	& -4.18	& 281.54	& -48.83\\[-1.2em]
2.27879	& 18.62 & 	-0.17	& 14.24	& 26.43	& 53.84	& 18.82	& 200.93	& -8.84	& -22.05	& 11.59\\[-1.2em]
2.28515	& 361.22 & 	-121.21	& 801.02	& 151.17	& 45.41	& -5.92	& 233.21	& -6.17	& 350.48	& -12.96\\[-1.2em]
2.28894	& 65.09 & 	-1.59	& 9.20	& -0.18	& 6.69	& 1.34	& 82.51	& 29.60	& 108.16	& -36.54\\[-1.2em]
2.29364	& 71.87 & 	-17.59	& 0.23	& 3.86	& -7.84	& -17.52	& -2.82	& -3.83	& 321.79& 	49.94\\[-1.2em]
2.29626	& 61.26 & 	-17.17	& 29.89	& 10.64	& 2.74	& 0.75	& 81.79	& 10.45	& 88.85	& -9.54\\[-1.2em]
\end{tabular}
\label{tab:EvenToOdd}
\end{table*}

Experimentally, the most easily accessible yellow exciton states are the 
odd-parity $\Gamma_4^-$ $P$-states. As the interseries dipole transition flips 
the parity, the coupled states will be green even-parity states with $S$-
and $D$-type envelopes. We now investigate two different scenarios. In the 
first, we select for the initial state the yellow $P$-exciton transforming like 
the basis state $z$ of the irreducible representation 
$\Gamma_4^-$ \cite{koster1963properties}. In the second scenario, we investigate 
the yellow $P$-exciton state transforming like the basis state $y$. 
In both cases, the photon polarization is along the $z$-direction.
From the product of the representations \cite{koster1963properties} $\Gamma_4^- 
\otimes \Gamma_4^- = \Gamma_1^+ + \Gamma_3^+ + \Gamma_4^+ + \Gamma_5^+$
we can determine which transitions to green states are allowed in principle.

We begin with the yellow $P$-exciton and the photon both transforming according 
to the $z$-component of the $\Gamma_4^-$ representation. This initial state can 
itself be excited using a one-photon absorption process with light polarized 
along the $z$-direction. Using the tables in Ref.~\cite{koster1963properties}, 
we can deduce that the corresponding green states transform according to 
$\Gamma_1^+$ and the $\psi^{3+}_1$-component of $\Gamma_3^+$. In 
Fig.~\ref{fig:OddToEvenZZ}, we show interseries transition
spectra in this configuration. We additionally list the results
for a selection of states in Table~\ref{tab:OddToEvenZZ}.

Using the Rydberg energies of the yellow and green exciton series, we can 
estimate which green principal quantum number belongs to states with maximum 
overlap with a yellow exciton state with given principal quantum number. 
In the following, we use the values
$E^\mathrm{y}_\mathrm{Ryd} = 86.04\,\mathrm{meV}$ \cite{Schoene2016} and
$E^\mathrm{g}_\mathrm{Ryd} = 150.4\,\mathrm{meV}$ \cite{Rommel2020Green}. The 
Bohr radii $a_0^\mathrm{y}$ and $a_0^\mathrm{g}$ are related to the Rydberg 
energies by
\begin{equation}
\frac{a_0^\mathrm{g}}{a_0^\mathrm{y}} \approx \frac{E^\mathrm{y}_\mathrm{Ryd}}{E^\mathrm{g}_\mathrm{Ryd}}\,.
\end{equation}
From a simple overlap argument, one would expect the transition strengths to be 
largest when initial and final state have comparable real-space extensions. 
As the linear extension of the excitons scales with the square of the 
principal quantum number $n$, we derive the estimate
\begin{equation}
n_\mathrm{g} = 
\sqrt{\frac{E^\mathrm{g}_\mathrm{Ryd}}{E^\mathrm{y}_\mathrm{Ryd}}} n_\mathrm{y} 
\approx 1.32\,n_\mathrm{y}\,,
\label{eq:nMaximumOverlap}
\end{equation}
to which the spectra in
Figs.~\ref{fig:OddToEvenZZ}$-$\ref{fig:EvenToOdd1SGreen}
fit approximately.

The resulting transition strengths are of the same order of magnitude as those 
found in Ref.~\cite{KruegerInterseries2019}. The strongest transition in 
Table~\ref{tab:OddToEvenZZ} is from the $2P$ yellow exciton to the exciton 
state with energy $E = 2.28895\,\mathrm{eV}$, which is a $3D$ state. The matrix 
elements become progressively weaker as the principal quantum number of the 
initial yellow state increases. At the same time, with increasing principal 
quantum number of the initial state, the green states with the highest 
transition strengths move to higher energies, in accordance with 
Eq.~\eqref{eq:nMaximumOverlap}.

Here, as well as in the following discussions, it is also important to remember 
that the choice of initial state does not only influence the strength of the 
transition, but also the energy gap between the states. This is most evident in 
the configuration in Sec.~\ref{subsec:EvenToOdd}, where the initial state is of 
even parity, leading to differences in the transition energies of up to 
$100\,\mathrm{meV}$.

We now proceed to the scenario that the $P$-exciton transforms according to the 
$\Gamma_4^-$ function $y$, meaning that the initial state can be excited using a 
single-photon absorption process with light polarized along the $y$-direction. 
Here, the corresponding green excitons transform like the $x$-component of $\Gamma_4^{+}$ and 
the $xz$-component of $\Gamma_5^{+}$ \cite{koster1963properties}. In Fig.~\ref{fig:OddToEvenYZ} 
we show a transition spectrum in this configuration. We additionally list the 
results for a selection of states in Table~\ref{tab:OddToEvenYZ}.

The strongest transition in Table~\ref{tab:OddToEvenYZ} is from the $2P$ yellow 
exciton to the exciton state with energy $E = 2.25655\,\mathrm{eV}$, which is 
the lowest lying $2S$ state. Nevertheless, it is hardly visible in our 
simulated spectrum in Fig.~\ref{fig:OddToEvenYZ} because
of its much larger width as compared with the other states.

As the principal quantum number of the initial yellow state increases, the 
matrix elements here also become progressively weaker. The region of green 
states with the strongest transition from a given yellow state does not obey
Eq.~\eqref{eq:nMaximumOverlap} as accurately as in the previous case, lying 
slightly lower energetically as expected. This could be related to the different
spatial extensions of the addressed green states, in addition to 
the generally approximate character of the overlap argument.

\subsection{Transitions from even parity yellow states to odd parity green states}
\label{subsec:EvenToOdd}

\begin{figure}
\includegraphics[width=\columnwidth]{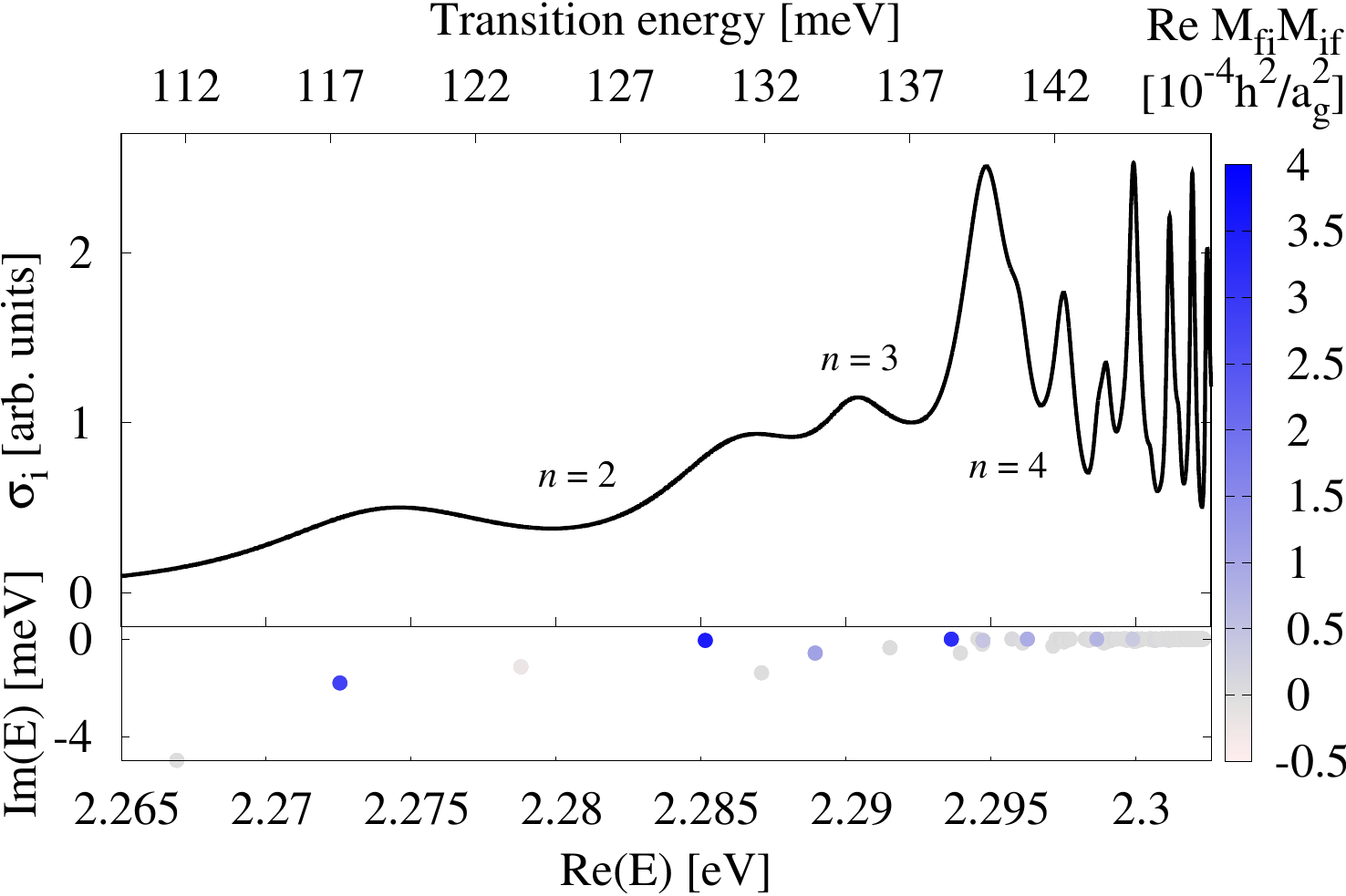}
\caption{Same as Fig.~\ref{fig:EvenToOdd}, but for the initial even parity green $1S$ $\Gamma_5^+$ exciton state transforming
as $xy$.}
\label{fig:EvenToOdd1SGreen}
\end{figure}

We finally investigate transitions from yellow even-parity states to green 
odd-parity states. The former can be excited using two-photon absorption 
processes. For these transitions, we have to consider states with 
irreducible representations appearing in the tensor product
$\Gamma_5^+ \otimes \Gamma_4^- = \Gamma_2^- + \Gamma_3^- + \Gamma_4^- + 
\Gamma_5^-$. In Fig.~\ref{fig:EvenToOdd}, we show spectra for transitions of 
this kind. As initial states, we chose excitons transforming according to the 
$xy$-component of the irreducible representation $\Gamma_5^+$. In 
Table~\ref{tab:EvenToOdd} we list the results for a selection of states.
In Fig.~\ref{fig:EvenToOdd1SGreen}, we additionally show the special case of 
the transitions where the initial state is the \emph{green} $1S$ state, which 
is energetically placed among the yellow excitons.

The strongest transition in Table~\ref{tab:EvenToOdd} is from the $2S$ yellow 
exciton to the green $2P$ exciton state with energy $E = 2.28515\,\mathrm{eV}$.
This is also the strongest transition we found among all configurations. 
This has to be balanced against the fact the the initial state is of even 
parity, which makes it inaccessible in one-photon transitions; it can, however, 
be excited using two-photon absorption.

We also investigated transitions from the $3D$ state, see panel (d) in 
Fig.~\ref{fig:EvenToOdd}. These seem to be substantially stronger than the 
transitions from the $3S$ states, but still weaker than those from the $1S$ and 
$2S$ excitons. Finally, there are several strong transitions starting from the 
green $1S$ exciton, but they are weaker than those from yellow $1S$ and $2S$ 
states.

\section{Conclusion and outlook}
\label{sec:Conclusion}
In this article, we have investigated interseries transitions between the 
yellow and green exciton series in the dipole approximation. We extended the 
calculations for the yellow-to-green interseries transitions performed in 
Ref.~\cite{KruegerInterseries2019} by including the complex valence 
band structure. To properly take into account the associated coupling between 
the green exciton states and the yellow continuum, we used the 
complex-coordinate rotation method for the calculation of the green exciton 
resonances as described in Ref.~\cite{Rommel2020Green}.

We considered different choices for the initial state in the spectral range of 
the yellow series, concentrating mostly on the odd-parity $P$-states, which are 
most easily accessible in one-photon absorption experiments. We distinguished
the cases where the photon that excites the initial exciton is polarized 
parallel to the photon affecting the interseries transition from the scenario 
in which they are orthogonally polarized. Additionally, we also calculated the 
probabilities for the transition from the even-parity yellow states to the
odd-parity green states, with the special case where the initial state is the 
\emph{green} $1S$-exciton.

The transition strengths are on the same order of magnitude in the different 
configurations, with those starting at an odd-parity yellow exciton being 
somewhat weaker than those starting at an even-parity yellow exciton. Of course,
the experimental preparation of the latter is more difficult, as a two-photon 
excitation is required. In all cases, increasing the principal quantum number 
of the initial state shifts the range of excited green states to higher 
energies, with an overall weakening of the transition strengths in most cases.

In this work, we use the dipole approximation, which is valid if the wavelength 
of the light affecting the interseries transition is much larger than the 
extension of the involved excitons. As shown in 
Ref.~\cite{KruegerInterseries2019}, this condition breaks down for transitions 
between the yellow and green series starting at $n \gtrsim 15$ for 
counter-propagating pump and probe beams. Extending our investigations to this 
parameter range thus requires going beyond the dipole approximation. 
Furthermore, an extension of our method to cover transitions between states of 
the yellow and blue series is relatively straightforward, {but} requires
the implementation of the conduction band Hamiltonian including the 
$\Gamma_8^-$ band.
Another possible route is to investigate the influence of an additional 
external field to fine tune the properties of the transitions.

{Finally, one of the aims of our manuscript was to provide theoretical predictions which can help
guide experimental investigations into the interseries transitions. 
While there has been some experimental work with respect to intraseries transitions within
the yellow series~\cite{Froehlich1985,Joerger2013}
and with respect to interseries transitions between the yellow and blue series~\cite{Schmutzler2013},
to the best of our knowledge, there have been no experimental studies into the yellow-to-green interseries transitions
investigated in this manuscript yet.
A comparison of our results with future experimental data is thus highly desirable.}

\acknowledgments
This work was supported by Deutsche Forschungsgemeinschaft (DFG)
through Grant No.~MA1639/13-1 and through Grant No.~SCHE 612/4-2 of the SPP 1929 `Giant Interactions in 
Rydberg Systems'.

\appendix

\section{The matrix element for $p_z$}
\label{sec:MatrixElementPz}
In the formalism of irreducible tensors, $p_z$ is given by
\begin{equation}
p_z = P^{(1)}_0 \,.
\end{equation}
In the supplemental material of Ref.~\cite{frankjanpolariton}, Eq.~(14)
provides the matrix elements for the operator
\begin{equation}
P^{(1)}_0 \left(I^{(1)} \cdot S^{(1)}_\mathrm{h}\right)\,,
\end{equation}
which we can use here.
Using the identity
\begin{align}
I^{(1)} \cdot S^{(1)}_\mathrm{h} = \frac{1}{2}(\boldsymbol{J}^2 - \boldsymbol{I}^2 - \boldsymbol{S}_\mathrm{h}^2) = \frac{2J(2J+2) - 11}{8}\,,
\end{align}
we can calculate the matrix element for $p_z$ using
\begin{align}
&\langle \Pi' |P^{(1)}_0| \Pi \rangle \\
&= \frac{8}{2J(2J+2) - 11} \left\langle \Pi'\left|P^{(1)}_0 \left(I^{(1)} \cdot S^{(1)}_\mathrm{h}\right)\right| \Pi \right\rangle\,.  \nonumber
\end{align}
Here, $|\Pi\rangle$ and $|\Pi'\rangle$ denote basis states as given in Eq.~\eqref{eq:basis}.

\section{Phonon-induced linewidths}
\label{sec:PhononLinewidths}
In order to use Eq.~\eqref{eq:phononlinewidth}, we need to determine 
the constant $\gamma^\mathrm{ph}_0$. According to Ref.~\cite{Grun1961}, the 
FWHM of the green $2P$-state at $T = 4\,\mathrm{K}$ is
$\gamma^{2\mathrm{P}} =17.7\,\mathrm{meV}$.
In Ref.~\cite{Rommel2020Green}, the complex coordinate rotation method was used 
to calculate the complex energies of the odd-parity green excitons, and to 
determine the linewidths $\gamma_\mathrm{cont}$
caused by the coupling of the green excitons to the yellow continuum. 
Here, we update this calculation by adding the Haken potential to the 
Hamiltonian,
and find $\gamma^{2\mathrm{P}}_\mathrm{cont} = 9.95\,\mathrm{meV}$ for the green $2P$ state.
We can thus estimate the phonon-induced linewidth of the $2P$ green 
exciton as
\begin{align}
\gamma_\mathrm{ph}(n=2) &= \gamma^{2\mathrm{P}} - 
\gamma^{2\mathrm{P}}_\mathrm{cont} \\
 &\approx 17.7\,\mathrm{meV} - 
9.95\,\mathrm{meV} = 7.05\,\mathrm{meV}\nonumber
\end{align}
leading to
\begin{align}
\gamma^\mathrm{ph}_0 &= 8\times\gamma_\mathrm{ph}(n=2)= 8\times 
7.05\,\mathrm{meV} = 56.4\,\mathrm{meV}\,.\label{eq:phononlinewidthvalue}
\end{align}

We associate to each resonance an effective quantum number $n_\mathrm{eff}$ as 
a function of the real part of the resonance energy $E$,
\begin{equation}
n_\mathrm{eff} = \sqrt{\frac{E_\mathrm{Ryd}}{E_\mathrm{gap} - E}} + \delta\,.
\end{equation}
The values $E_\mathrm{Ryd} = 142\,\mathrm{meV}$, $E_\mathrm{gap} = 
2.30292\,\mathrm{eV}$ and $\delta = 0.1$ were obtained by a phenomenological 
fit to the odd-parity green excitons in an updated version of the calculation in 
Ref.~\cite{Rommel2020Green}, where we included the Haken potential. Note that 
these values should not be taken as the literal Rydberg energy and quantum 
defect, as the inclusion of the Haken potential distorts the Rydberg 
spectrum.


%

\end{document}